\documentclass[fleqn,usenatbib]{mnras}

\usepackage{newtxtext,newtxmath}

\usepackage[T1]{fontenc}
\usepackage{ae,aecompl}

\usepackage{graphicx}	
\usepackage{amsmath}	
\usepackage{amssymb}	
\usepackage{longtable}

\title[H\,I Asymmetries in LVHIS, VIVA and HALOGAS Galaxies]{H\,I Asymmetries in LVHIS, VIVA and HALOGAS Galaxies.}

\author[T.N.~Reynolds, et al.]{T.N.~Reynolds$^{1,2,3}$\thanks{tristan.reynolds@research.uwa.edu.au}, 
T.~Westmeier$^{1,3}$,
L.~Staveley-Smith$^{1,3}$,
G.~Chauhan$^{1,3}$,
\newauthor C.D.P.~Lagos$^{1,3}$\\
\\
$^1$International Centre for Radio Astronomy Research (ICRAR), The University of Western Australia,\\ 35 Stirling Hwy, Crawley, WA, 6009, Australia\\
$^2$CSIRO Astronomy and Space Science, Australia Telescope National Facility, P.O. Box 76, Epping NSW 1710, Australia\\
$^3$ARC Centre of Excellence for All Sky Astrophysics in 3 Dimensions (ASTRO 3D)
}

\date{Accepted 2020 February 25. Received 2020 February 18; in original form 2019 November 14}

\pubyear{2020}

\begin{document}
\label{firstpage}
\pagerange{\pageref{firstpage}--\pageref{lastpage}}
\maketitle

\begin{abstract}
We present an analysis of morphological, kinematic and spectral asymmetries in observations of atomic neutral hydrogen (H\,\textsc{i}) gas from the Local Volume H\,\textsc{i} Survey (LVHIS), the VLA Imaging of Virgo in Atomic Gas (VIVA) survey and the Hydrogen Accretion in Local Galaxies Survey (HALOGAS). With the aim of investigating the impact of the local environment density and stellar mass on the measured H\,\textsc{i} asymmetries in future large H\,\textsc{i} surveys, we provide recommendations for the most meaningful measures of asymmetry for use in future analysis. After controlling for stellar mass, we find signs of statistically significant trends of increasing asymmetries with local density. The most significant trend we measure is for the normalised flipped spectrum residual ($A_{\mathrm{spec}}$), with mean LVHIS and VIVA values of $0.204\pm0.011$ and $0.615\pm0.068$ at average weighted $10^{\mathrm{th}}$ nearest-neighbour galaxy number densities of $\log(\rho_{10}/\mathrm{Mpc}^{-3})=-1.64$ and 0.88, respectively. Looking ahead to the WALLABY survey on the Australian Square Kilometre Array Pathfinder (ASKAP), we estimate that the number of detections will be sufficient to provide coverage over 5 orders of magnitude in both local density and stellar mass increasing the dynamic range and accuracy with which we can probe the effect of these properties on the asymmetry in the distribution of atomic gas in galaxies.
\end{abstract}

\begin{keywords}
galaxies: groups: general -- galaxies: clusters: general -- radio lines: galaxies
\end{keywords}



\section{INTRODUCTION}
\label{sec:intro}

The stellar and gaseous (atomic hydrogen, H\,\textsc{i}) disks of galaxies are found to have morphologies which vary from symmetric to highly asymmetric \citep[first studied in H\,\textsc{i} by][]{Baldwin1980}. The review by \cite{Jog2009} indicates that spiral galaxies commonly exhibit morphological asymmetries in their H\,\textsc{i}. The gravitational potential from the baryonic (stellar, gas, dust) and non-baryonic (dark matter) matter is the dominant driver of galaxy morphology and kinematics. In an isolated, massive system, the morphology and kinematics are expected to be symmetrical around the galaxy's centre. Perturbations away from a symmetrical system are then expected to be due to external influences of the environment \citep[e.g.][]{Hibbard2001}, although isolated galaxies are also observed with asymmetric H\,\textsc{i} morphologies \citep[e.g.][]{Portas2011,Athanassoula2010}. Proposed mechanisms that can create asymmetries in isolated galaxies include gas accretion along filaments \citep[e.g.][]{Bournaud2005b,Mapelli2008}, minor mergers of satellites \citep[e.g.][]{Zaritsky1997,Bournaud2005b,LagosP2018} and fly-by interactions \citep[e.g.][]{Mapelli2008}. H\,\textsc{i} is a sensitive probe of environmental effects as H\,\textsc{i} is easily observable at larger radii than the stellar component and will be the first to exhibit signs of external influences \citep[e.g.][]{Giovanelli1985,Solanes2001,Rasmussen2006,Rasmussen2012,Westmeier2011,Denes2014,Odekon2016}. Proposed external mechanisms for causing asymmetries in a galaxy's observed H\,\textsc{i} include ram pressure stripping \citep[e.g.][]{Gunn1972,Kenney2004} by dense intergalactic medium (IGM), high relative velocity galaxy interactions \citep[harassment,][]{Moore1996,Moore1998}, low relative velocity galaxy interactions \citep[tidal stripping,][]{Moore1999,Koribalski2009,English2010}, galaxy mergers \citep[e.g.][]{Zaritsky1997,Rubin1970} and asymmetric gas accretion \citep[e.g.][]{Bournaud2005b,Sancisi2008,LagosP2018}. Ram pressure stripping in particular is proposed to be the dominant driver of the evolution of galaxy morphology in clusters \citep[e.g.][]{Boselli2006}.

Early work measured H\,\textsc{i} asymmetries in integrated spectra, as these are more easily obtained for large samples of galaxies compared with spatially resolved H\,\textsc{i} observations, finding asymmetric fractions of $\gtrsim50\%$ based on the the flux ratio asymmetry, $A_{\mathrm{flux}}$, \citep[the ratio of the integrated flux of the two halves of the spectrum divided at the systemic velocity,][]{Richter1994,Haynes1998,Matthews1998}. \cite{Espada2011} used the AMIGA \citep[Analysis of the interstellar Medium in Isolated GAlaxies,][]{VerdesMontenegro2005} sample of isolated galaxies to quantify the intrinsic scatter in observed $A_{\mathrm{flux}}$ values, finding the distribution could be described as a half Gaussian centred on 1 (perfectly symmetric) with a $1\sigma$ standard deviation of 0.13. Within the tail of the Gaussian distribution, $9\%$ and $2\%$ of the AMIGA sample were found to have $A_{\mathrm{flux}}>2\sigma$ and $3\sigma$, respectively \citep{Espada2011}. Applying the $2\sigma$ cut to the isolated samples of \cite{Haynes1998} and \cite{Matthews1998} gives fractions of $9\%$ and $17\%$, respectively. Recent studies of single dish spectra of Virgo and Abell\,1367 cluster galaxies considered $A_{\mathrm{flux}}>2\sigma$ or $3\sigma$ as likely produced by external, environmental influences and found asymmetric fractions of $\sim16$--$26\%$ \citep{Scott2018} and $27\%$ for close galaxy pairs \citep{Bok2018} using the AMIGA sample $3\sigma$ significance threshold. \cite{Watts2020} classified xGASS \citep{Catinella2018} galaxies as satellites or centrals and measured $A_{\mathrm{flux}}$, finding that satellites show a higher frequency of asymmetries than centrals, which supports the previous findings of asymmetries being more common in higher density environments. However, taking this further and understanding the physical origin of spectral asymmetries requires knowledge of the spatial distribution of gas within galaxies and the gas kinematics. 

Spatially resolved asymmetry analyses were first carried out on near-infrared (near-IR) galaxy images using Fourier analysis which found $\sim30\%$ to show signs of morphological asymmetries \citep{Rix1995,Zaritsky1997}. A decade later \cite{Bournaud2005b} found a significantly higher asymmetric fraction ($\sim60\%$) also using Fourier analysis of near-IR images. In the following years, Fourier analysis was applied to H\,\textsc{i} integrated intensity images, obtaining asymmetric fractions of $\sim17$--$27\%$ in the Eridanus and Ursa Major galaxy groups \citep{Angiras2006,Angiras2007} and $\sim30\%$ from a sample of WHISP\footnote{The Westerbork HI Survey of Irregular and Spiral Galaxies \citep{Swaters2002}} galaxies \citep{vanEymeren2011b}. The H\,\textsc{i} studies have the advantage of probing out to larger radii than the near-IR image analysis. Focusing on the gas kinematics, \cite{Swaters1999} showed that differences in the rotation curves of a galaxy's approaching and receding sides is a sign of kinematic asymmetry for two WHISP spiral galaxies. \cite{vanEymeren2011a} measured the kinematic lopsidedness for a larger sample of 70 WHISP galaxies, concluding that this parameter can be an over- or underestimate if local distortions are present in the disk and cannot be used alone to characterise two dimensional asymmetry.

There are currently no large statistical samples of spatially resolved galaxies in H\,\textsc{i} that can match sample sizes of all-sky single dish surveys. These all-sky coverage H\,\textsc{i} surveys \citep[e.g. HIPASS and ALFALFA,][respectively]{Barnes2001,Haynes2018} have limited spatial resolution. This is about to change with the advent of new radio interferometers including the Australian Square Kilometre Array Pathfinder \citep[ASKAP,][]{Johnston2008}, the APERture Tile in Focus upgrade to the Westerbork Synthesis Telescope \citep[APERTIF,][]{Verheijen2008} and the Karoo Array Telescope \citep[MeerKAT,][]{Jonas2016}. ASKAP, which is fitted with phased array feed receivers \citep{DeBoer2009,Hampson2012,Hotan2014,Schinckel2016}, has a wide field of view giving it fast survey speed capabilities while retaining the increased resolution of an interferometer. The Widefield ASKAP L-band Legacy All-sky Blind Survey \citep[WALLABY,][Koribalski et al. in prep.]{Koribalski2012} will use the increased survey speed and high resolution to detect H\,\textsc{i} emission in $\sim500\,000$ galaxies across $\sim75\%$ of the sky \citep{Duffy2012}. Several thousand of these detections, including all the HIPASS sources, will be spatially resolved. This will provide the largest environmentally unbiased sample for which morphological and kinematic asymmetries can be measured. The rest of the WALLABY detections will be limited to investigating asymmetries in their integrated spectra. However, the significantly smaller synthesised beam of WALLABY compared to HIPASS and ALFALFA ($0.5\arcmin$ vs $15.5\arcmin$ and $3.5\arcmin$, respectively) will result in a lower fraction of confused detections, increasing the number of spectra uncontaminated by near neighbours (e.g. from galaxies in pairs and groups).

In preparation for WALLABY it is useful to determine the optimal measures of H\,\textsc{i} asymmetry to parameterise the galaxies which will be detected. To do this we only consider existing surveys which spatially resolve galaxies in H\,\textsc{i}, as single dish surveys (e.g. HIPASS and ALFALFA) cannot be used to measure morphological or kinematic asymmetries. Existing publicly available interferometric surveys include THINGS \citep{Walter2008}, Little THINGS \citep{Hunter2012}, LVHIS \citep{Koribalski2018}, VIVA \citep{Chung2009}, HALOGAS \citep{Heald2011} and ATLAS$^{3\mathrm{D}}$ \citep{Cappellari2011}. The spectral and angular resolution and spectral line sensitivity of LVHIS is similar to that achievable with WALLABY \citep{Koribalski2018} and is a perfect candidate for testing asymmetry measures. However, LVHIS only contains 82 galaxies and probes isolated galaxy and group environments. We require additional surveys probing higher densities with comparable spectral resolution, physically resolved spatial scale, stellar masses, and sensitivity to study the effect of environment on measured asymmetry in late-type galaxies. These surveys must also lie within an optical background survey footprint \citep[e.g. SDSS or 6dFGS,][respectively]{Strauss2002,Jones2009} to derive environment densities consistently across each H\,\textsc{i} survey. VIVA, which probes the significantly higher densities of the Virgo cluster, satisfies these criteria (Table~\ref{table:obs_params}), except for only covering the high stellar masses in LVHIS (Fig.~\ref{fig:optical_hists}). To control for stellar mass, we include HALOGAS, which is also well matched to LVHIS and VIVA (Table~\ref{table:obs_params}), and probes similar densities to LVHIS (Fig.~\ref{fig:density_vs_distance}). Neither THINGS nor Little THINGS span the full mass range of LVHIS and VIVA. THINGS covers the high masses of VIVA \citep{deblok2008} and LITTLE THINGS covers the low masses of the majority of LVHIS \citep{Oh2015}. ATLAS$^{3\mathrm{D}}$ is a survey of early type galaxies and has a spectral resolution 2 to 4 times lower than LVHIS, VIVA or HALOGAS \citep{Serra2012}. Our final galaxy samples include galaxies from the LVHIS, VIVA and HALOGAS surveys.

In this work we aim to determine the best asymmetry parameters for quantifying H\,\textsc{i} asymmetries to use on data to be produced by future large surveys (e.g. WALLABY) for investigating the influence of environment density and stellar mass on H\,\textsc{i} asymmetry parameters. This paper is structured as follows. In Section~\ref{sec:data} we introduce the surveys used in this work and describe our analysis in Section~\ref{sec:analysis}. We present our discussion and conclusions in Sections~\ref{sec:discussion} and \ref{sec:conclusion}, respectively. Throughout, we use velocities in the optical convention ($cz$) and the heliocentric reference frame, adopting a flat $\Lambda$CDM cosmology using $H_0=67.7$, concordant with \textit{Planck} \citep{Planck2016}.

\section{SAMPLE SELECTION}
\label{sec:data}

We use the Local Volume H\,\textsc{i} Survey \citep[LVHIS,][]{Koribalski2018}, the VLA Imaging of Virgo in Atomic Gas survey \citep[VIVA,][]{Chung2009} and the Hydrogen Accretion in Local Galaxies survey \citep[HALOGAS,][]{Heald2011} to investigate the influence of environment density on measured morphological, kinematic and spectral asymmetry parameters. Here we provide a brief overview of each survey and direct the reader to the survey papers for more details. We present a summary of the survey parameters in Table~\ref{table:obs_params} and the stellar and H\,\textsc{i} mass distributions, $B-$band magnitude, specific star formation rate ($\mathrm{sSFR}=\mathrm{SFR/M_*}$), H\,\textsc{i} detection signal to noise ratio (SNR) and the H\,\textsc{i} diameter in Fig.~\ref{fig:optical_hists} (panels ordered from upper left to lower right). We calculate the H\,\textsc{i} masses for LVHIS, VIVA and HALOGAS using a simplified form of Equ.~50 from \cite{Meyer2017},
\begin{equation}
	\displaystyle \frac{M_{\mathrm{H\,\textsc{i}}}}{\mathrm{M}_{\odot}} \sim \frac{2.35\times10^5}{1+z}\left(\frac{D}{\mathrm{Mpc}}\right)^2\left(\frac{S_{\mathrm{int}}}{\mathrm{Jy\,km\,s}^{-1}}\right),
	\label{equ:mhi}
\end{equation}
where $S_{\mathrm{int}}$ is the integrated flux and $D$ is the galaxy distance. LVHIS star formation rates (SFRs) are computed in \cite{Shao2018} using \textit{IRAS} 60 and $100\,\mu$m fluxes, which for consistency we use to compute the HALOGAS and VIVA SFRs following equations~2--5 from \cite{Kewley2002}. We note that the LVHIS dwarf galaxy SFRs will be lower estimates due to lower dust opacity \citep[e.g.][]{Shao2018}. Dwarf galaxy SFRs are more accurately derived by including UV and H\,$\alpha$ luminosities \citep[e.g.][and references therein]{Cluver2017}, however we do not have UV and H\,$\alpha$ luminosities for all of the three samples to consistently derive SFRs. For LVHIS and VIVA, H\,\textsc{i} properties are listed in the overview papers \citep[][respectively]{Koribalski2018,Chung2009}. However for consistency among the three surveys, we perform our own source finding on the survey H\,\textsc{i} spectral line cubes (Section~\ref{sec:analysis}) and re-measure H\,\textsc{i} properties from the detected sources. We find good agreement between the published properties and our values.

\begin{table}
	\centering
    \caption{Survey parameters summary. The galaxy sub-samples used in this work were selected to lie within the footprints of the 6dFGS (LVHIS) and SDSS (VIVA and HALOGAS) optical redshift catalogues and have publicly available H\,\textsc{i} spectral line cubes. The number of galaxies for each survey sub-sample that are resolved by $\geq3$ beams is indicated in brackets. The $3\sigma$ column density sensitivity is for a channel width of 10\,km\,s$^{-1}$.}
	\label{table:obs_params}
	\begin{tabular}{lccr}
		\hline
		Survey & LVHIS & VIVA & HALOGAS \\ \hline
		Telescope & ATCA & VLA & WSRT \\
		Total Galaxies [$N$] & 82 & 53 & 24 \\
		Galaxy Sub-samples [$N$] & 73 (61) & 45 (41) & 18 (18) \\
		Distance [Mpc] & $<10$ & $\sim16.5$ & $<25$ \\
        $3\sigma$ Column Density & & & \\ 
        Sensitivity [$10^{19}\mathrm{cm}^{-2}$]  & 0.4--7.4 & 0.9--9.0 & 0.2--0.6 \\
        Stellar Mass [$\log(M_{*}/\mathrm{M}_{\odot})$] & 6--11 & 9--11 & 7.5--11 \\
        \\
        \textbf{Resolution} &  &  & \\
        Spectral [km\,s$^{-1}$] & 4--8 & 10 & 10 \\
		Angular [arcsec] & $\gtrsim40$ & 15 & 40 \\
		Physical [kpc] & 0.6--2.3  & 1.5 & 0.8--4.8 \\ \hline
	\end{tabular}
\end{table}

\subsection{LVHIS}
\label{s-sec:lvhis}

The Local Volume H\,\textsc{i} Survey \citep[LVHIS,][]{Koribalski2018} was carried out on the Australia Telescope Compact Array (ATCA), observing H\,\textsc{i} in 82 nearby ($<10$\,Mpc) gas-rich spiral, dwarf and irregular galaxies. The majority of LVHIS galaxies are members of local groups or pairs and have at least one close neighbour with an angular separation of $<300\arcsec$ (e.g. projected separation $<17$\,kpc) and a systemic velocity $<800$\,km\,s$^{-1}$ \citep[group membership and number of close neighbours are tabulated in][]{Koribalski2018}. All galaxies were observed using a minimum of three ATCA configurations to provide good uv coverage and sensitivity to H\,\textsc{i} gas on different spatial scales. We exclude galaxies which are not covered by the 6dF Galaxy Survey due to incompleteness (see Section~\ref{s-sec:environment}). This reduces our sample to 73 LVHIS galaxies. We use LVHIS H\,\textsc{i} spectral line cubes made using natural weighting.\footnote{LVHIS H\,\textsc{i} spectral line cubes with natural weighting are publicly available for download at \url{http://www.atnf.csiro.au/research/LVHIS/LVHIS-database.html}.} The LVHIS H\,\textsc{i} spectral line cubes have a 4\,km\,s$^{-1}$ spectral resolution\footnote{Two LVHIS galaxies have a spectral resolution of 8\,km\,s$^{-1}$ (LVHIS\,004 and LVHIS\,005) and one has a spectral resolution of 2\,km\,s$^{-1}$ (LVHIS\,079).}, $5\arcsec\times5\arcsec$ pixels and a synthesised beam of $\gtrsim40\arcsec\times40\arcsec$. We use LVHIS stellar masses from \cite{Wang2017}.

\subsection{VIVA}
\label{s-sec:viva}

The VLA Imaging of Virgo in Atomic Gas survey \citep[VIVA,][]{Chung2009} imaged 53 late-type galaxies in H\,\textsc{i} using the Karl G. Jansky Very Large Array (VLA) at an angular resolution of $\sim15$\arcsec. Here we use a subsample of 45 galaxies for which the reduced H\,\textsc{i} spectral line cubes are available.\footnote{VIVA H\,\textsc{i} spectral line cubes are publicly available for download at \url{http://www.astro.yale.edu/cgi-bin/viva/observations.cgi}.} The VIVA H\,\textsc{i} spectral line cubes have a 10\,km\,s$^{-1}$ spectral resolution and $5\arcsec\times5\arcsec$ pixels. We calculate VIVA stellar masses using the empirical relation from \cite{Taylor2011},
\begin{multline}
	\displaystyle \log(M_*/\mathrm{M}_{\odot}) = a + b\,(g-i) - 0.4m + 0.4D_{\mathrm{mod}} + 0.4M_{\mathrm{sol}} \\  - \log(1+z) - 2\log(h/0.7),
	\label{equ:mstar}
\end{multline}
where $a,b=-1.197,1.431$ are empirically determined constants based on the chosen magnitude and colour from \cite{Zibetti2009}, we use the SDSS $g-i$ colour, $m$ is the $g-$band apparent magnitude, $D_{\mathrm{mod}}$ is the distance modulus and $M_{\mathrm{sol}}=5.11$ is the absolute magnitude of the sun in the $g-$band \citep{Willmer2018}. SDSS magnitudes have been corrected for foreground Galactic extinction \citep{Kim2014}. All VIVA physical parameters are calculated assuming these galaxies are located at the distance of the Virgo cluster (16.5\,Mpc).

\subsection{HALOGAS}
\label{s-sec:halogas}

The Hydrogen Accretion in Local Galaxies survey \citep[HALOGAS,][]{Heald2011} on the Westerbork Synthesis Radio Telescope (WSRT) imaged 24 spiral galaxies in H\,\textsc{i} at an angular resolution of $\sim40$\arcsec. The HALOGAS sample is composed of galaxies residing in pairs and groups and includes both dominant and minor group members \citep[group membership is tabulated in][]{Heald2011}. Here we use a subsample of 18 galaxies\footnote{HALOGAS H\,\textsc{i} spectral line cubes are publicly available for download at \url{http://www.astron.nl/halogas/data.php}.} which lie within the SDSS spectroscopic survey footprint \citep{Strauss2002}. Only 17 are specific HALOGAS targets, with the final galaxy, NGC\,2537, lying within the same observed field as UGC\,4278. We use the low-resolution H\,\textsc{i} spectral line cubes, which have a physical resolution that is comparable to the LVHIS and VIVA observations. The HALOGAS H\,\textsc{i} spectral line cubes have a $\sim5$\,km\,s$^{-1}$ spectral resolution and $5\arcsec\times5\arcsec$ pixels. We also calculate the HALOGAS stellar masses using Equ.~\ref{equ:mstar}, replacing the $g-i$ colour with the $B-V$ colour and the $g-$band apparent magnitude with the 2MASS $J-$band apparent magnitude. The constants for the $B-V$ colour and $J-$band magnitude are $a,b=-1.135,1.267$ and the 2MASS absolute $J-$band magnitude of the sun $M_{\mathrm{sol}}=4.54$ \citep[][respectively]{Zibetti2009,Willmer2018}.

\begin{figure*}
	\centering
	\includegraphics[width=16cm]{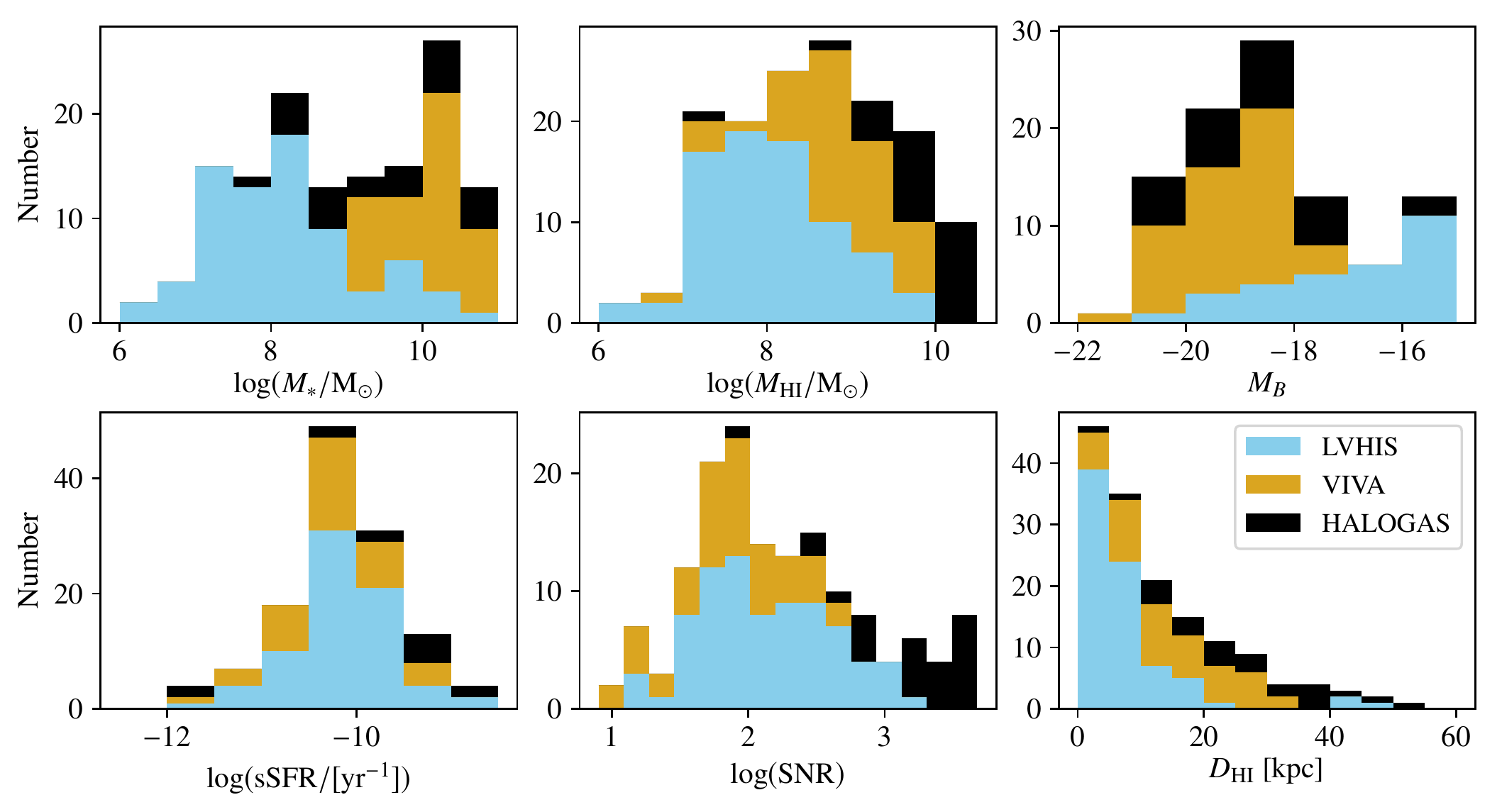}
		\caption{Stacked histograms of LVHIS (blue), VIVA (yellow) and HALOGAS (black) stellar and H\,\textsc{i} masses, 25$^{\mathrm{th}}$ mag\,arcsec$^{-2}$ $B-$band magnitude, specific star formation rate ($\mathrm{sSFR}=\mathrm{SFR/M_*}$), the integrated H\,\textsc{i} detection signal to noise ratio (SNR) and the H\,\textsc{i} diameter  (upper left, upper centre, upper right, lower left, lower centre and lower right panels, respectively). $B-$band magnitudes and optical diameters are taken from the LVHIS, VIVA and HALOGAS overview papers \citep[][respectively]{Koribalski2018,Chung2009,Heald2011}. LVHIS stellar masses are from \citet{Wang2017} and the VIVA and HALOGAS stellar mass calculates are described in Sections~\ref{s-sec:viva} and \ref{s-sec:halogas}. SoFiA returns the integrated H\,\textsc{i} SNR, which we correct for the beam solid angle, and the H\,\textsc{i} diameter, which is computed by fitting an ellipse to the moment map, converted from an angular to physical size.}
	\label{fig:optical_hists}
\end{figure*}

\subsection{Environment}
\label{s-sec:environment}

We determine the local environment galaxy number density for the galaxy samples using a 3D friends of friends (FOF) approach for the $K^{\mathrm{th}}$ nearest neighbour (NN) method cross matching with optical redshift survey catalogues. We use the 6dF Galaxy Survey \citep[6dFGS,][]{Jones2009} for LVHIS, the Extended Virgo Cluster Catalogue \citep[EVCC,][]{Kim2014} for VIVA and the catalogue of Sloan Digital Sky Survey (SDSS) sources with spectroscopic redshifts \citep{Strauss2002} for HALOGAS. The EVCC contains a subset of galaxies from the SDSS spectroscopic catalogue which are determined to be members of the Virgo cluster. For computing VIVA densities, we assume a distance of 16.5\,Mpc to the Virgo cluster \citep{Mei2007} and give each VIVA galaxy a random distance selected from a Gaussian distribution centred on 16.5\,Mpc with a standard deviation of 1.72\,Mpc, the virial radius of the Virgo cluster \citep{Hoffman1980}. All densities calculated and discussed in this work are galaxy number densities (i.e. the number of galaxies per Mpc$^3$), not to be confused with the density of the IGM or intercluster medium (ICM), which is not easily measurable.

We calculate the weighted density to the $N^{\mathrm{th}}$ nearest neighbour following the Bayesian metric of \cite{Cowan2008},
\begin{equation}
	\displaystyle \rho_N=\frac{C_N}{\sum_{i=1}^N d_i^3},
	\label{equ:nn_density}
\end{equation}
where $C_N$ is a constant, derived empirically, such that the mean density agrees with the true density calculated on a uniform density, regular grid (e.g. for $N=10$ nearest neighbours, $C_{10}=11.48$) and $\sum_{i=1}^N d_i^3$ is the sum of the cubed luminosity distances, $d$, to the $N$ nearest neighbours. Equ.~\ref{equ:nn_density} is sensitive to the distances to all $N$ nearest neighbours and provides a better estimate of the local galaxy density than using only the $N^{\mathrm{th}}$ nearest neighbour. Take as an example, the weighted number density of two galaxies for $N=10$ where one galaxy has the first nine neighbours within 0.5\,Mpc and the $10^{\mathrm{th}}$ nearest galaxy is at 3\,Mpc and the other galaxy only has two neighbours within 0.5\,Mpc with the other eight neighbours at 2.5--3\,Mpc. The $10^{\mathrm{th}}$ nearest galaxy is at 3\,Mpc in both cases, hence the density to the $10^{\mathrm{th}}$ nearest neighbour will be the same as this ignores the inner nine neighbours. The weighted nearest neighbour on the other hand is sensitive to the distances to all 10 neighbours. It   will return a higher number density for the first galaxy with more weighting going to the inner nine galaxies compared with the second galaxy which will return a lower density with more weighting going to the more distant neighbours. The measured densities for both galaxies will be higher than simply using the $10^{\mathrm{th}}$ nearest neighbour distance as the use of the inner nine neighbours in the calculation will more accurately describe the local density.

\begin{figure}
	\centering
	\includegraphics[width=\columnwidth]{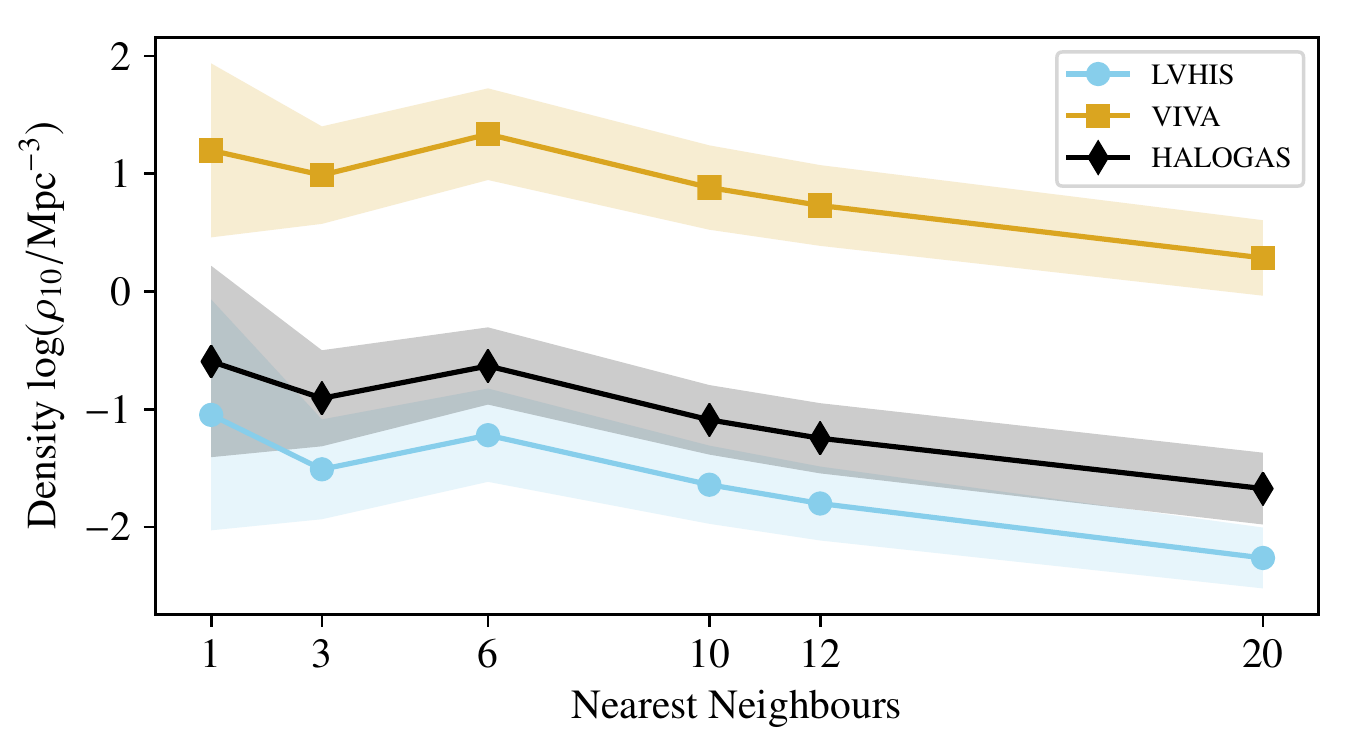}
		\caption{Mean environment density vs number of nearest neighbours used for calculation for LVHIS, VIVA and HALOGAS (blue circle, yellow square and black diamond, respectively). The shaded region indicates the 1$\sigma$ dispersion in the calculated density distributions.}
	\label{fig:density_vs_nn}
\end{figure}

\begin{figure}
	\centering
	\includegraphics[width=\columnwidth]{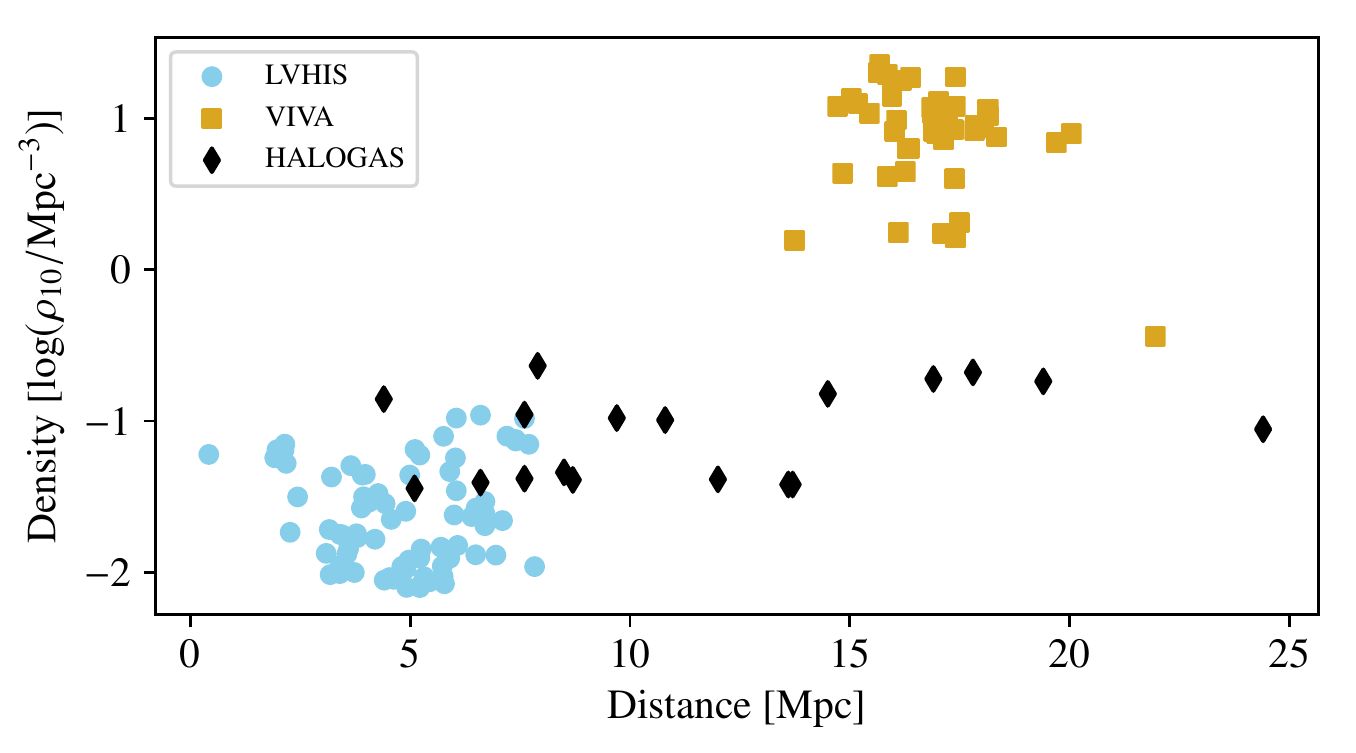}
		\caption{Environment density vs physical distance. We show the weighted 10$^{\mathrm{th}}$ nearest neighbour densities calculated using Equ.~\ref{equ:nn_density} after correcting for survey completeness as described in Section~\ref{s-sec:environment}. Distances for LVHIS and HALOGAS are taken from \citet{Koribalski2018} and \citet{Heald2011}, respectively. For VIVA we assume a distance of 16.5\,Mpc to the Virgo cluster and give each VIVA galaxy a random distance selected from a Gaussian distribution centred on 16.5\,Mpc with a standard deviation of 1.72\,Mpc, the distance to and virial radius of the Virgo cluster virial radius \citep[][respectively]{Mei2007,Hoffman1980}.}
	\label{fig:density_vs_distance}
\end{figure}

We calculate luminosity distances from galaxy redshifts using the \textsc{python} module \textsc{astropy} function \textsc{luminosity\_distance}. We tested measuring density to the 1$^{\mathrm{st}}$, 3$^{\mathrm{rd}}$, 6$^{\mathrm{th}}$, 10$^{\mathrm{th}}$, 12$^{\mathrm{th}}$ and 20$^{\mathrm{th}}$ nearest neighbours. In each regime the relative difference in mean environment density remains nearly constant (see Fig.~\ref{fig:density_vs_nn}), which is a good indication that VIVA is probing a different density environment than LVHIS and HALOGAS. The only exception to this is for the 1$^{\mathrm{st}}$ nearest neighbour, which also has the largest dispersion due to the random nature of the proximity of the nearest galaxy and hence does not provide a good indication of the local environment (e.g. field, group or cluster galaxy number densities). We find using the 10 nearest neighbours \citep[i.e. the same number of neighbours used by][]{Cowan2008} produces well-defined density distributions with the dispersion in density for each survey $\sim1$\,dex. Using less than 10 neighbours produces larger dispersions in densities and using more than 10 neighbours results in densities that are reduced by an order of magnitude compared to the local density (Fig.~\ref{fig:density_vs_nn}). We show the environment density (corrected for observational biases) distributions against the physical distances of each galaxy in Fig.~\ref{fig:density_vs_distance}. The mean densities are $\log(\rho_{10}/\mathrm{Mpc}^{-3})=-1.64,\,0.88\,\mathrm{and}\,-1.09$ for LVHIS, VIVA and HALOGAS, respectively. Throughout this work we refer to LVHIS and HALOGAS as low density relative to the high density Virgo cluster VIVA galaxies.

Both 6dFGS and SDSS have observational magnitude limits, resulting in fainter galaxies falling below the detection limit with increasing distance. We correct for the bias this creates in the computed densities using a mock survey catalogue for which we compute the measured density for the full mock and after applying the magnitude limits for each survey (2MASS $H=12.95$, $J=13.75$ and $K=12.65$ for 6dFGS and SDSS $r=17.77$ for SDSS). The mock survey is created using one of the simulation boxes from the Synthetic UniveRses For Surveys \citep[SURFS,][]{Elahi2018} $N$-body simulations denoted medi-SURFS (210\,cMpc/h on a side and $1536^3$ dark matter particles). The dark matter halos in this $N$-body simulation are populated with galaxies using the semi-analytic model \textsc{shark} \citep{Lagos2018}. We then create a lightcone from the simulation box using the \textsc{stingray} code (Obreschkow, in prep.), which uses an extension of the algorithm described by \cite{Blaizot2005} and \cite{Obreschkow2009}. We use the spectral energy distribution modelling codes \textsc{prospect}\footnote{\url{https://github.com/asgr/ProSpect}} (Robotham et al., submitted) and \textsc{viperfish}\footnote{\url{https://github.com/asgr/Viperfish}} \citep{Lagos2019} to derive magnitudes for the mock galaxies in the SDSS $r$--band filter and the VISTA $J$--, $H$-- and $K$--band filters. \textsc{shark} reproduces extremely well the near-IR number counts reported by \cite{Driver2016} and the luminosity functions at different redshifts, which are shown by \cite{Lagos2019}. This gives us confidence that we can use \textsc{shark}-based lightcones to inform us about necessary corrections to our environmental density calculation. We convert the 2MASS $J$--, $H$-- and $K$--band magnitude limits used in 6dFGS to VISTA bands using the calibration from \cite{GonzalezFernandez2018}. Medi-SURFS has a mass resolution limit of $\sim10^8\,\mathrm{M}_{\odot}$ (dark matter particle mass resolution: $4.13\times10^7\,\mathrm{M}_{\odot}$/h). We apply this mass cut to the SDSS and 6dFGS catalogues to consistently calculate the density correction factors based on the medi-SURFS mock catalogue. We calculate SDSS stellar masses using Equ.~\ref{equ:mstar} with SDSS $g-$ and $i-$band magnitudes and 6dFGS stellar masses using equation~1 from \cite{Beutler2013}, $\log(M_*)=\log(0.48-0.59C_{b_J-r_F}) + \log(M_J^{\mathrm{sun}}-M_J)/2.5$, where $C_{b_J-r_F}$ is the 2MASS $b_J-r_F$ colour, $M_J$ is the $J-$band absolute magnitude and $M_J^{\mathrm{sun}}=3.7$ is the $J-$band absolute magnitude of the sun \citep{Worthey1994}. 

\begin{figure}
	\centering
	\includegraphics[width=\columnwidth]{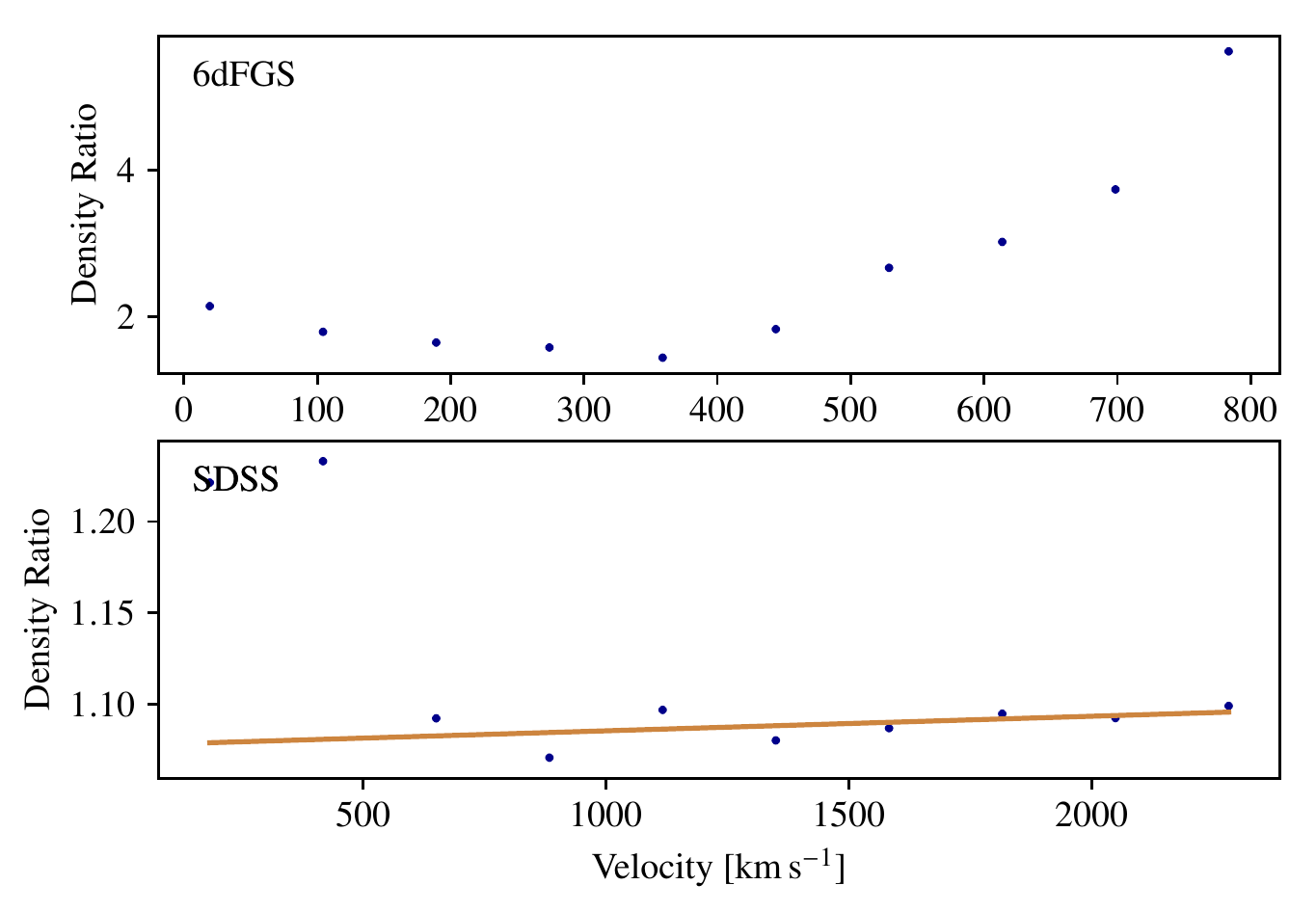}
		\caption{Variation of the ratio between the expected 10$^{\mathrm{th}}$ nearest neighbour environment density for the full mock catalogue and the 10$^{\mathrm{th}}$ nearest neighbour environment density computed after applying the optical survey flux limits with velocity ($cz$) for 6dFGS and SDSS (upper and lower panels, respectively). The line in the lower panel is an exponential fit to the binned data. We exclude the first two bins from the fit as these bins are dominated by small numbers compared to the other bins (e.g. $\sim20$ vs $\gtrsim200$ galaxies per bin). We only perform a fit for correcting SDSS as 6dFGS has the additional issue of sky completeness, which is direction dependent.}
	\label{fig:density_ratio_vs_velocity}
\end{figure}

For SDSS we compute the local densities for every galaxy in the mock, bin the galaxies in redshift and take the ratio of the average density in each bin for the full mock to the flux limited mock (lower panel of Fig.~\ref{fig:density_ratio_vs_velocity}). We exclude the first two bins from the fit as these bins are dominated by small numbers compared to the other bins (e.g. $\sim20$ vs $\gtrsim200$ galaxies per bin). We then fit an exponential function to the binned ratios, which we use to scale the computed HALOGAS densities. For VIVA, we use the mean ratio (1.08) at the distance of the Virgo cluster, 16.5\,Mpc ($\sim1140$\,km\,s$^{-1}$), to scale all VIVA densities. We find good agreement between our corrected densities in the Virgo cluster with the computed densities for ATLAS$^{3\mathrm{D}}$ Virgo cluster galaxies \citep{Serra2012}. Correcting for 6dFGS incompleteness is more complicated due to incomplete sky coverage \citep[e.g. due to the Galactic plane and fibre breakages,][]{Jones2004}. We account for this by applying a sky completeness mask after the flux limits. We cannot calculate the scaling ratio in redshift bins as we did for SDSS since the sky completeness of 6dFGS is direction-dependent. However, we show the effect of the 6dFGS flux limits in the top panel of Fig.~\ref{fig:density_ratio_vs_velocity}, which illustrates the greater effect the flux limits have on the 6dFGS compared to SDSS computed densities. Instead, we rotate the mock catalogue with respect to the sky completeness mask and LVHIS galaxy sky positions and compute the mean density for each LVHIS galaxy from 50 different orientations. We then take the ratio of the full mock to the flux and completeness limited mock and scale each LVHIS galaxy by the computed ratio for the given galaxy.

\section{ANALYSIS}
\label{sec:analysis}

We use the Source Finding Application\footnote{We use SoFiA v1.2.1, which can be found at \url{https://github.com/SoFiA-Admin/SoFiA/}.} \citep[SoFiA,][]{Serra2015a} to extract H\,\textsc{i} emission $>3.5\sigma$ from the LVHIS, VIVA and HALOGAS cubes with default parameters for the Smooth$+$Clip (S$+$C) finder. The S$+$C finder smooths the cube over a number of spatial and spectral scales and identifies emission above the defined threshold which are merged at the end. We set minimum merging sizes of 5 pixels in both spatial and spectral dimensions and have reliability set to $>0.7$. SoFiA then creates integrated intensity (moment 0) and velocity field (moment 1) maps, integrated spectra and detected source properties.

\subsection{Spectral Asymmetries}
\label{s-sec:spec_asyms}

\begin{figure}
	\centering
	\includegraphics[width=\columnwidth]{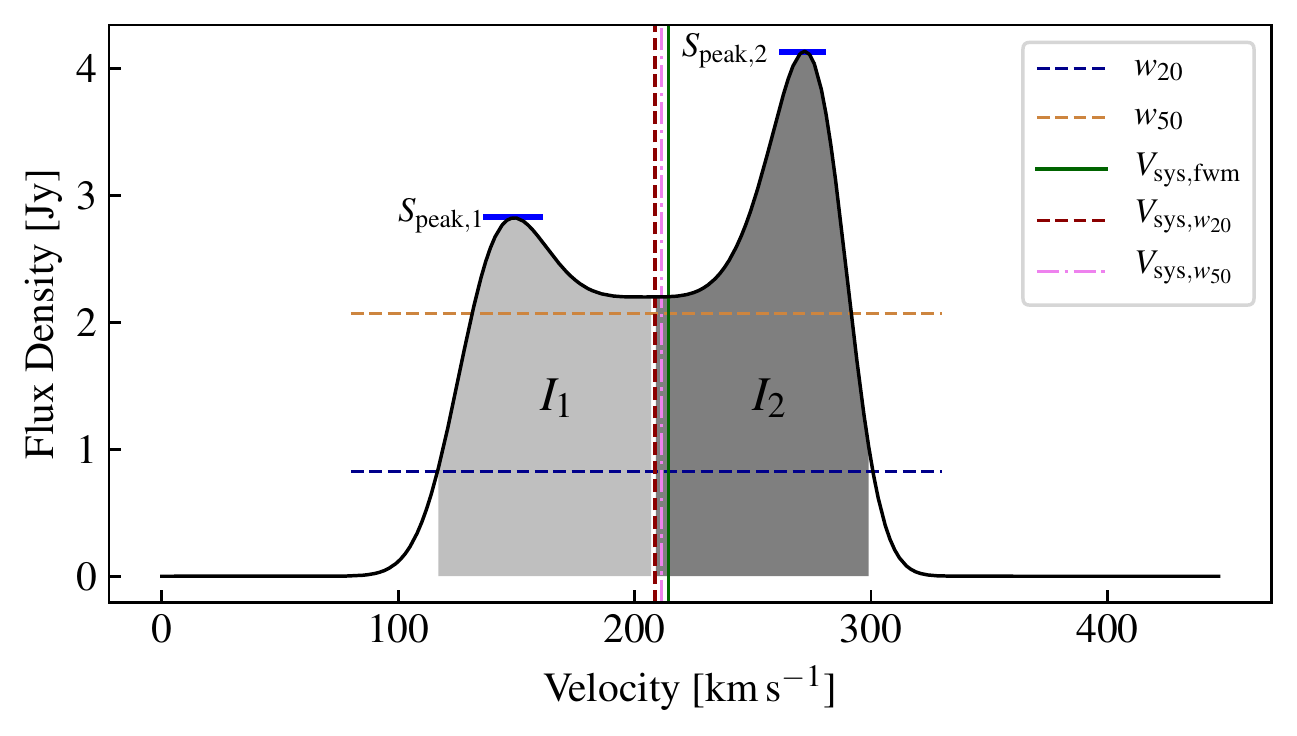}
		\caption{Example of an asymmetric H\,\textsc{i} spectrum indicating the line widths, $w_{20}=184$\,km\,s$^{-1}$ and $w_{50}=160$\,km\,s$^{-1}$ (dashed blue and orange horizontal lines, respectively), the flux weighted mean systemic velocity, $v_{\mathrm{sys,fwm}}=214$\,km\,s$^{-1}$, the systemic velocity defined as the midpoint of the spectrum at $20\%$ the spectrum's peak height, $V_{\mathrm{sys},w_{20}}=209$\,km\,s$^{-1}$ and the systemic velocity defined as the midpoint of the spectrum at $50\%$ the spectrum's peak height, $V_{\mathrm{sys},w_{50}}=211$\,km\,s$^{-1}$ (solid green, dashed red and dot-dashed violet vertical lines, respectively). Also indicated are the peak fluxes of the lower and upper wings of the spectrum ($S_{\mathrm{peak,1}}$ and $S_{\mathrm{peak,2}}$, respectively) used to calculate $A_{\mathrm{peak}}$ and the light and dark shaded regions integrated to calculate $A_{\mathrm{flux}}$ ($I_1$ and $I_2$, respectively). The spectral asymmetries for this spectrum are $\Delta V_{\mathrm{sys}}=3$\,km\,s$^{-1}$, $A_{\mathrm{spec}}=0.15$, $A_{\mathrm{flux}}=1.09$ and $A_{\mathrm{peak}}=1.47$.}
	\label{fig:example_spectrum}
\end{figure}

The integrated spectrum of a galaxy provides global H\,\textsc{i} information on a galaxy as a whole compared with spatially resolved maps, which provide more localised information of the distribution of H\,\textsc{i} within a galaxy. Most galaxies that have been observed in H\,\textsc{i} and will be observed by future surveys will be spatially unresolved, limiting studies of asymmetries in these galaxies to their integrated spectra. We present an example mock spectrum illustrating the spectral asymmetries in Fig.~\ref{fig:example_spectrum}. We find spectral resolutions of 4--10\,km\,s$^{-1}$ are sufficient to measure the spectral asymmetries described in this section.

\subsubsection{Velocity Differences}
\label{ss-sec:vel_diff}

The H\,\textsc{i} systemic velocity can be measured in different ways. In this work we measure the systemic velocity using two definitions: the flux weighted mean systemic velocity, $V_{\mathrm{sys,fwm}}$, and the systemic velocity defined as the midpoint of the spectrum at $50\%$ of the spectrum's peak height (i.e. based on the $w_{50}$ line width), $V_{\mathrm{sys},w_{50}}$. We then compute the offset between these two H\,\textsc{i} systemic velocity definitions,
\begin{equation}
	\displaystyle \Delta V_{\mathrm{sys}}=|V_{\mathrm{sys,fwm}}-V_{\mathrm{sys},w_{50}}|.
	\label{equ:vrad_diff}
\end{equation}
We calculate $w_{50}$ using the peak of the full spectrum, rather than using the $50\%$ point from the peaks of each wing of the spectrum, for consistency as not all spectra show a double horned shape and will only have a single peak from which to measure the $50\%$ flux value. In general this will reduce the measured offsets as using the $50\%$ flux level from two peaks will give systemic velocities shifted towards the lower peak wing of the spectrum compared with using only the peak of the full spectrum. By definition the integrated flux in each half of the spectrum divided by $V_{\mathrm{sys,fwm}}$ is equal, while $V_{\mathrm{sys},w_{50}}$ depends on the spectrum's peak and slope of the outer edges. Hence, $\Delta V_{\mathrm{sys}}$ is sensitive to differences in the H\,\textsc{i} distribution in the spectrum as $V_{\mathrm{sys},w_{50}}$ can be offset from $V_{\mathrm{sys,fwm}}$ depending on the shape of the spectrum. 

A galaxy's H\,\textsc{i} gas is more easily affected by gravitational and hydrodynamical interactions (e.g. tidal and ram pressure stripping) compared with its stellar component, which can result in an offset between the systemic velocities measured from optical and H\,\textsc{i} surveys. We calculate the difference between the flux weighted mean systemic velocity from the H\,\textsc{i} spectra, $V_{\mathrm{sys,fwm}}$, and the optical velocity, $V_{\mathrm{opt}}$, obtained from 6dFGS (LVHIS) or SDSS (VIVA and HALOGAS),
\begin{equation}
	\displaystyle \Delta V_{\mathrm{sys,opt}}=|V_{\mathrm{sys,fwm}}-V_{\mathrm{opt}}|.
	\label{equ:vopt_diff}
\end{equation}
The 6dFGS and SDSS optical velocities have uncertainties of $\sim46$ and 30\,km\,s$^{-1}$, respectively \citep{Jones2009,Strauss2002}.

\subsubsection{Flux Ratio Asymmetry}
\label{ss-sec:flux_asym}

The simplest and most commonly used measure of H\,\textsc{i} spectral asymmetry is defined as the ratio of the integrated flux in the left and right halves of the spectrum divided at the systemic velocity \citep[e.g.][]{Richter1994,Haynes1998,Espada2011,Scott2018},
\begin{equation}
	\displaystyle A_{\mathrm{flux}}=\frac{I_1}{I_2}=\frac{\int_{v_{\mathrm{low}}}^{v_{\mathrm{sys},w_{20}}} Idv}{\int_{v_{\mathrm{sys},w_{20}}}^{v_{\mathrm{high}}} Idv},
	\label{equ:flux_asym}
\end{equation}
where $I_{1}$ and $I_{2}$ are the integrated fluxes in the lower and upper halves of the spectrum (shaded regions in Fig.~\ref{fig:example_spectrum}) integrated from $v_{\mathrm{low}}=v_{\mathrm{sys},w_{20}}-w_{20}/2$ to $v_{\mathrm{sys},w_{20}}$ and from $v_{\mathrm{sys},w_{20}}$ to $v_{\mathrm{high}}=v_{\mathrm{sys},w_{20}}+w_{20}/2$, respectively. $v_{\mathrm{sys},w_{20}}$ is the systemic velocity defined as the midpoint of the spectrum at the $20\%$ flux level (the $w_{20}$ line width) and $v_{\mathrm{low}}$ and $v_{\mathrm{high}}$ are the velocities at which the flux density drops to $20\%$ of the peak flux density. For channels bridging the edges of regions $I_{1}$ and $I_{2}$, the flux is assigned to $I_{1}$ or $I_{2}$ based on the fraction of the channel within either region. For example, a channel centred on $v_{\mathrm{sys},w_{20}}$ will be evenly divided between $I_{1}$ and $I_{2}$. Similarly, only half the flux in a channel centred on $v_{\mathrm{low}}$ or $v_{\mathrm{high}}$ will contribute to $I_{1}$ and $I_{2}$, respectively). If the flux ratio $A_{\mathrm{flux}}<1$ then the inverse is taken so that $A_{\mathrm{flux}}\geq1$. A perfectly symmetric spectrum has $A_{\mathrm{flux}}=1$, with larger values indicating the spectrum is more asymmetric. 

\subsubsection{Peak Flux Ratio Asymmetry}
\label{ss-sec:peak_asym}

A related asymmetry parameter to $A_{\mathrm{flux}}$, is taking the ratio between the left and right peaks of the spectrum \citep[e.g. the 'height asymmetry index',][]{Matthews1998}, 
\begin{equation}
	\displaystyle A_{\mathrm{peak}}=\frac{S_{\mathrm{peak,1}}}{S_{\mathrm{peak,2}}},
	\label{equ:peak_asym}
\end{equation}
where $S_{\mathrm{peak,1}}$ and $S_{\mathrm{peak,2}}$ are the peak fluxes of the lower and upper wings of the spectrum (illustrated in Fig.~\ref{fig:example_spectrum}). As with $A_{\mathrm{flux}}$, if $A_{\mathrm{peak}}<1$ we take the inverse such that $A_{\mathrm{peak}}\geq1$. We note this is the inverse of the \cite{Matthews1998} definition. The peak flux ratio is limited to use on double horn profiles with defined peaks. $A_{\mathrm{peak}}$ is sensitive to peaks in the noise and will not be reliably measurable for noisy spectra. This does not affect our spectra which generally have high signal to noise ratios of $\mathrm{SNR}>50$.

\subsubsection{Flipped Spectrum Residual}
\label{ss-sec:spec_res}

We also look for signs of asymmetry in the residual of the integrated spectrum, which has not been investigated in previous studies. We define the spectrum residual as the sum of the absolute differences between the flux in each channel of the spectrum and the flux in the spectral channel of the spectrum flipped about the systemic velocity normalised by the integrated flux of the spectrum,
\begin{equation}
	\displaystyle A_{\mathrm{spec}}=\frac{\sum_{i}|S(i)-S_{\mathrm{flip}}(i)|}{\sum_{i}|S(i)|},
	\label{equ:spec_res}
\end{equation}
where $S(i)$ and $S_{\mathrm{flip}}(i)$ are the fluxes in channel $i$ of the original and flipped spectrum, respectively. Here we use the flux weighted mean systemic velocity, $V_{\mathrm{sys,fwm}}$, so that the spectrum is flipped around the centre of mass so that $A_{\mathrm{spec}}$ is sensitive to differences in the spectral shape (e.g. higher flux peak or more extended, lower flux on one side of the spectrum).

\subsection{Spatially Resolved Asymmetries}
\label{s-sec:morph_lop}

Spatially resolved galaxies provide the additional information of the distribution of H\,\textsc{i} gas within the galaxies, down to the physical scale resolved, and the gas kinematics. Morphological asymmetries use the two-dimensional information of the integrated intensity (moment 0) map. The gas kinematics uses the full three-dimensional information of the relative positions of the gas across the plane of the galaxy observed at different frequencies/velocities. There are a number of morphological parameters that are computed for optical images of galaxies, including the concentration \citep{Bershady2000}, asymmetry \citep{Abraham1996,Conselice2000}, smoothness/clumpiness \citep{Conselice2003}, $M_{20}$ \citep{Lotz2004} and Gini \citep{Abraham2003}, which have been previously applied to H\,\textsc{i} moment 0 maps \citep[e.g.][]{Holwerda2011a,Holwerda2011b,Holwerda2011c,Holwerda2011d}. \cite{Giese2016} found that the signal to noise ratio (SNR) affects the reliability of the optical parameters when applied to H\,\textsc{i} maps, with low SNR data not producing meaningful results. Accounting for SNR, \cite{Giese2016} found the \cite{Conselice2000} asymmetry to be the only parameter to provide useful information for measuring deviations away from a symmetrical disk. Hence from the optical morphology parameters we only consider the \cite{Conselice2000} asymmetry. In this work we only measure morphological and kinematic asymmetries for galaxies resolved by $\geq3$ beams, as \cite{Giese2016} found that a galaxy needs to be resolved by three or more beams to distinguish between differing levels of asymmetry.

\subsubsection{Moment 0 Asymmetry}
\label{ss-sec:mom0_asym}

\begin{figure}
	\centering
	\includegraphics[width=\columnwidth]{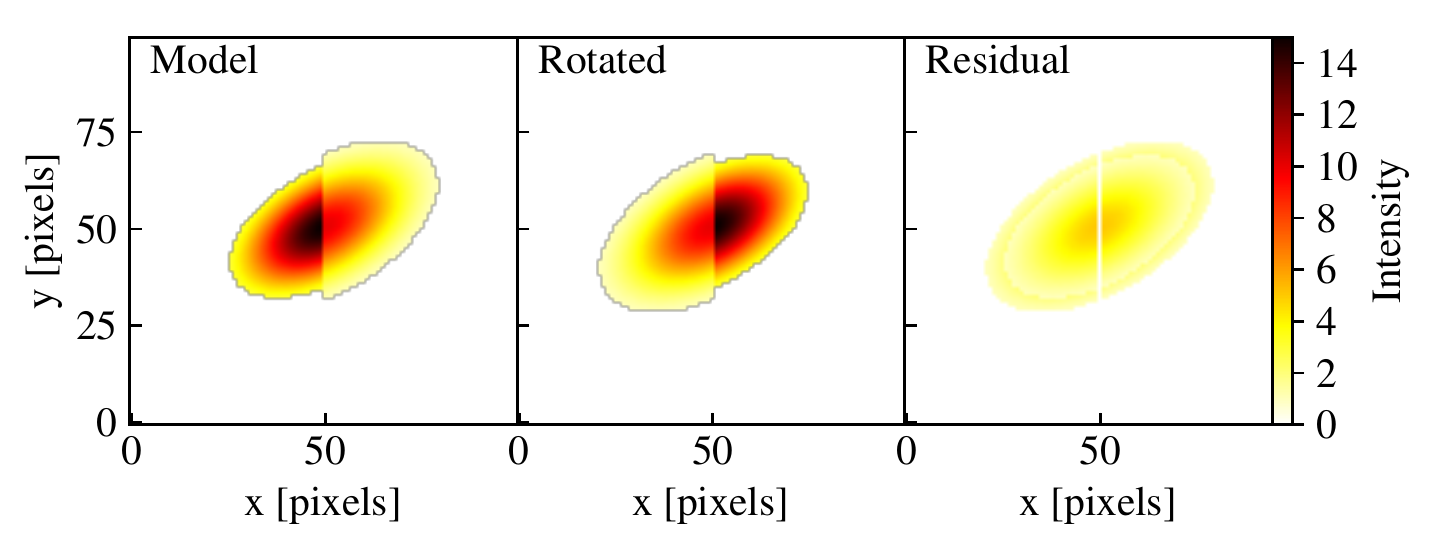}
		\caption{Example of an asymmetric integrated intensity (moment 0) map demonstrating the moment 0 map asymmetry parameter, $A_{\mathrm{map}}$. The left, centre and right panels show the model galaxy, model rotated by 180$^{\circ}$ and absolute residuals, respectively. $A_{\mathrm{map}}$ is then calculated by summing the absolute residuals. For this model $A_{\mathrm{map}}=0.23$.}
	\label{fig:example_map}
\end{figure}

We use the asymmetry parameter, $A$, \citep[first used by][]{Abraham1996} used for classification in optical studies to quantify the morphological asymmetry in the moment 0 map using the more recent definition of \cite{Conselice2000}, which includes a term to correct for bias due to noise and background,
\begin{equation}
	\displaystyle A_{\mathrm{map}}=\frac{\sum_{i,j}|I(i,j)-I_{180}(i,j)|}{2\sum_{i,j}|I(i,j)|}-\frac{\sum_{i,j}|B(i,j)-B_{180}(i,j)|}{2\sum_{i,j}|I(i,j)|},
	\label{equ:mom0_asym_corr}
\end{equation}
where $I(i,j)$ and $I_{180}(i,j)$ are the integrated intensity in pixel $(i,j)$ in the moment 0 map and moment 0 map rotated by $180^{\circ}$, respectively. Similarly, $B(i,j)$ and $B_{180}(i,j)$ are the integrated intensity in pixel $(i,j)$ of the bias image and the rotated bias image. The bias image is used to account for the effect of noise in the moment 0 map. We rotate the map about the flux centre of mass of the galaxy. Optical studies \citep[e.g.][]{Lotz2004} generally minimise $A_{\mathrm{map}}$ by shifting the centre of rotation around the central pixels due to the 5--10 times higher physical scale resolution of optical images compared with the H\,\textsc{i} map physical resolutions (e.g. $\sim0.02$--0.1\,kpc vs $\sim0.6$--4.8\,kpc) and is not required for H\,\textsc{i} maps. We create the bias image by placing the SoFiA source mask in a section of the spectral cube not containing any H\,\textsc{i} signal and integrating the signal in each voxel in the mask to create a moment 0 image. \cite{Giese2016} showed that the bias-corrected asymmetry, $A_{\mathrm{map}}$, provides a lower limit on the intrinsic asymmetry and that the asymmetry can be better determined using machine learning and a library of model galaxies, however this is beyond the scope of the current work. We illustrate the calculation for a noiseless model galaxy, the model rotated by $180^{\circ}$ and the absolute residuals which are then summed (left, centre and right panels of Fig.~\ref{fig:example_map}, respectively).

\subsubsection{Fourier Analysis}
\label{ss-sec:fourier}

We use a Fourier analysis to compute morphological lopsidedness in the moment 0 maps by decomposing the maps into their Fourier modes as performed on near-infrared (near-IR) images \citep[e.g.][]{Zaritsky1997,Bournaud2005b} and on H\,\textsc{i} moment 0 maps \citep{Angiras2006,Angiras2007,vanEymeren2011b},
\begin{equation}
	\displaystyle \sigma(r,\phi)=a_0(r)+\sum_n a_n(r)\cos[n(\phi-\phi_n(r))],
	\label{equ:mom0_fd}
\end{equation}
where $a_0(r)$ is the mean density at radius $r$, $a_n(r)$ is the $n^{\mathrm{th}}$ harmonic coefficient at $r$, $\phi_n$ is the phase of the $n^{\mathrm{th}}$ coefficient and $\phi$ is the azimuthal angle in the plane of the galaxy. The lopsidedness is then calculated as the ratio of the $n=1$ harmonic coefficient, $a_1$, to the mean density, $a_0$, at each radius,
\begin{equation}
	\displaystyle A_1(r)=\frac{a_1(r)}{a_0(r)}.
	\label{equ:fd_asym}
\end{equation}
We use the \textsc{idl} code \textsc{kinemetry} \citep{Krajnovic2006} to Fourier decompose the moment 0 maps and determine the values $a_0$ and $a_1$ in annuli of width half the synthesised beam. This results in correlation between adjacent rings, but ensures we are not under sampling.

\cite{vanEymeren2011b} calculated the mean value, $\langle A_1\rangle$, of the inner and outer H\,\textsc{i} disk, defined as $r/R_{25}<1$ and $r/R_{25}>1$, respectively, where $R_{25}$ is the optical radius. The optical radius is $\sim4$--5 times the scale length used for Fourier analysis in the near-IR \citep{vanderKruit1982,Zaritsky1997,Bournaud2005b}, thus the Fourier analysis of H\,\textsc{i} maps probes lopsidedness to significantly larger radii than possible in the near-IR. We follow \cite{vanEymeren2011b} and compute $\langle A_1\rangle$ for the inner and outer disks, $\langle A_{1,r/R_{25}<1}\rangle$ and $\langle A_{1,r/R_{25}>1}\rangle$, respectively.

\subsubsection{Velocity Map Weighted Median Absolute Deviation}
\label{ss-sec:vel_mad}

Distortions and asymmetries in a galaxy's velocity field can also be quantified by taking the sum of the velocity at each pixel in the moment 1 map and in the map rotated by 180$^{\circ}$. This is similar to the morphological asymmetry described in Section~\ref{ss-sec:mom0_asym}, however, instead of summing the residuals between the original and rotated maps, we take the weighted median absolute deviation (WMAD) of the summed velocity map, $V_{\mathrm{WMAD}}$. We weight by the square root of the flux in each pixel of the moment 0 map as a proxy for the signal to noise ratio (SNR). Weighting by the flux predominantly reduces the contribution of the outer regions of the disk, with less H\,\textsc{i} emission and lower SNR, which can dominate the calculated WMAD as we found for the VIVA galaxies by comprising a larger fraction of the pixels in the map. Hence the flux weighted MAD is biased towards asymmetries in the inner regions and $V_{\mathrm{WMAD}}$ is less sensitive to asymmetries at the galaxy edges. 

\begin{figure}
	\centering
	\includegraphics[width=\columnwidth]{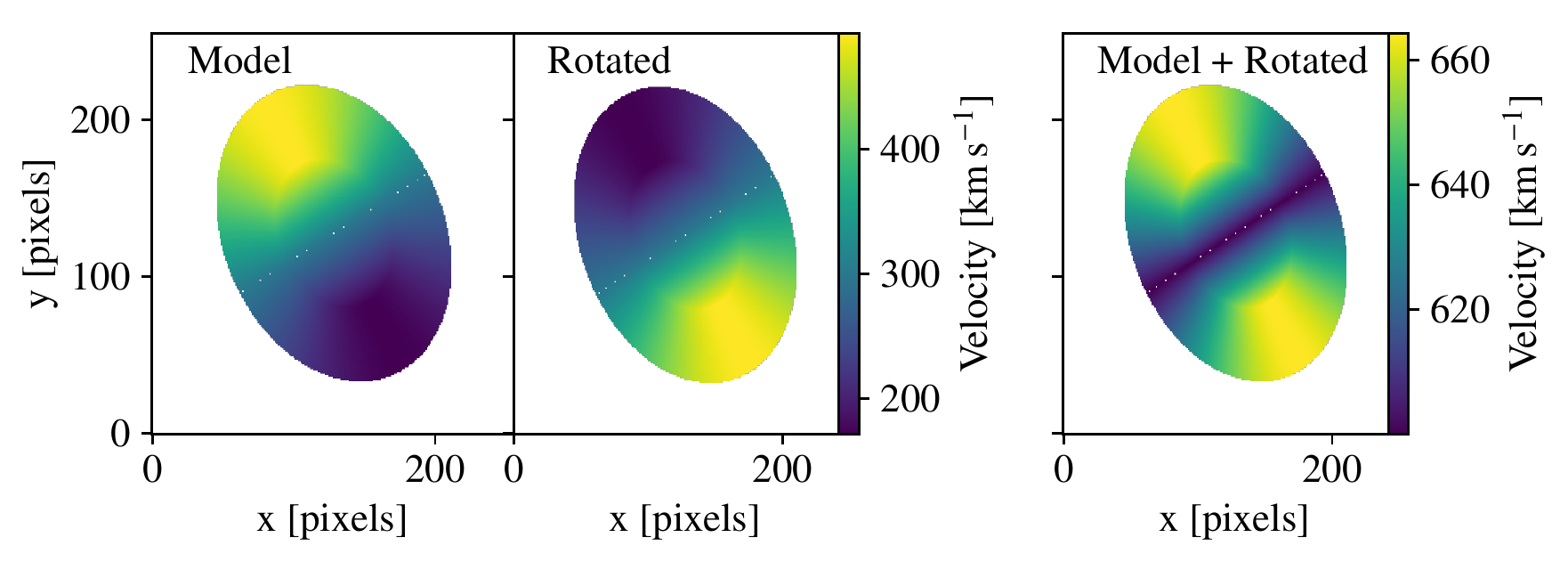}
		\caption{Example of an asymmetric velocity field (moment 1) map demonstrating the moment 1 map weighted median absolute deviation parameter, $V_{\mathrm{WMAD}}$. The left, centre and right panels show the model galaxy, the model rotated by 180$^{\circ}$ and the sum of the first two panels, respectively. $V_{\mathrm{WMAD}}$ is the weighted median absolute deviation of the right panel. For this model $V_{\mathrm{WMAD}}=18$\,km\,s$^{-1}$. Scaling $V_{\mathrm{WMAD}}$ by the 95$^{\mathrm{th}}$ percentile of the difference map gives $A_{\mathrm{vel}}=0.057$.}
	\label{fig:example_velmadev}
\end{figure}

A perfectly symmetric galaxy will have $V_{\mathrm{WMAD}}=0$\,km\,s$^{-1}$ while larger values of $V_{\mathrm{WMAD}}$ indicate increasing levels of asymmetry. To illustrate the computation of $V_{\mathrm{WMAD}}$, we show an example velocity field for a noiseless model galaxy, the model rotated by $180^{\circ}$ and the sum of the model and rotated model in Fig.~\ref{fig:example_velmadev} (left, centre and right panels, respectively). $V_{\mathrm{WMAD}}$ has an inherent bias due to velocity field deviations of the same magnitude in galaxies with large rotational velocities (i.e. greater dynamic range in the summed map) producing larger $V_{\mathrm{WMAD}}$ values compared to galaxies with small rotational velocities. We account for this bias by scaling $V_{\mathrm{WMAD}}$ by the 95$^{\mathrm{th}}$ percentile of the difference map between the original and rotated maps, which we use as a proxy for the maximum rotational velocity, producing a unitless velocity asymmetry parameter $A_{\mathrm{vel}}$,
\begin{equation}
	\displaystyle A_{\mathrm{vel}}=\frac{\mathrm{median}(|I_{i,j}(V+V_{180})_{i,j}-\langle V+V_{180}\rangle|)}{(V-V_{180})_{95^{\mathrm{th}}\,\mathrm{percentile}}},
	\label{equ:vel_wmad}
\end{equation}
where $I_{i,j}$ is the flux in pixel ($i,j$), $(V+V_{180})_{i,j}$ is the summed velocity in pixel ($i,j$), $\langle V+V_{180}\rangle$ is the mean velocity of the summed velocity field and $(V-V_{180})_{95^{\mathrm{th}}\,\mathrm{percentile}}$ is the $95^{\mathrm{th}}$\,percentile of the residual velocity field. $A_{\mathrm{vel}}$ has no dependence on the systemic velocity of the galaxy.

\begin{table}
	\centering
    \caption{Mean and standard deviation of $A_{\mathrm{vel}}$ calculated from rotating about the centre of mass (COM) or the kinematic centre (KIN) derived by fitting a tilted ring model to the velocity field with \textsc{3dbarolo}.}
	\label{table:wmad_params}
	\begin{tabular}{lcr}
		\hline
		Survey & $A_{\mathrm{vel,COM}}$ & $A_{\mathrm{vel,KIN}}$ \\ \hline
		LVHIS   & $0.063\pm0.049$ & $0.054\pm0.043$ \\
		VIVA    & $0.075\pm0.056$ & $0.076\pm0.063$ \\
		HALOGAS & $0.037\pm0.020$ & $0.032\pm0.018$ \\ \hline
	\end{tabular}
\end{table}

The galaxy centre about which the velocity field is rotated can be defined from either the integrated intensity map, the centre of mass (COM), or the velocity field map, the kinematic centre, which are not guaranteed to be the same position. We tested rotating the galaxy about the COM vs the kinematic centre determined from a tilted ring fit to the velocity field map using \textsc{3dbarolo}. We choose to use the COM galaxy centre as both the COM and kinematic centres produced similar $A_{\mathrm{vel}}$ values and this allows us to use additional galaxies for which the tilted ring fits failed. There is some scatter in the $A_{\mathrm{vel}}$ values derived from each centre of rotation, however the means of each sample are the same within the standard deviations (Table~\ref{table:wmad_params}). 

We note that for our analysis of $A_{\mathrm{vel}}$ we exclude galaxies with H\,\textsc{i} diameters $D_{\mathrm{HI}}<5$\,kpc, which removes galaxies with low rotational velocities. For these galaxies, the 95$^{\mathrm{th}}$ percentile of the difference map will not necessarily trace the rotational velocity due to non-negligible random motions within the galaxies. The majority of galaxies with $D_{\mathrm{HI}}<5$\,kpc are LVHIS galaxies (Fig.~\ref{fig:optical_hists} bottom, right panel). We find including these galaxies raises the mean LVHIS $A_{\mathrm{vel}}$ value by biasing the mean LVHIS asymmetry towards small galaxies with higher $A_{\mathrm{vel}}$ values as $\sim50\%$ of the LVHIS galaxies have $D_{\mathrm{HI}}<5$\,kpc. Removing these galaxies produces comparable galaxy samples in low and high density environments with the rotational velocity influence removed. An advantage of larger galaxy samples with future surveys will be the ability to select subsamples containing equal galaxy numbers covering the same physical sizes across different environments.

\section{DISCUSSION}
\label{sec:discussion}

\subsection{Asymmetry vs Environment Density}
\label{s-sec:asym_env}

\begin{figure*}
	\centering
	\includegraphics[width=17.7cm]{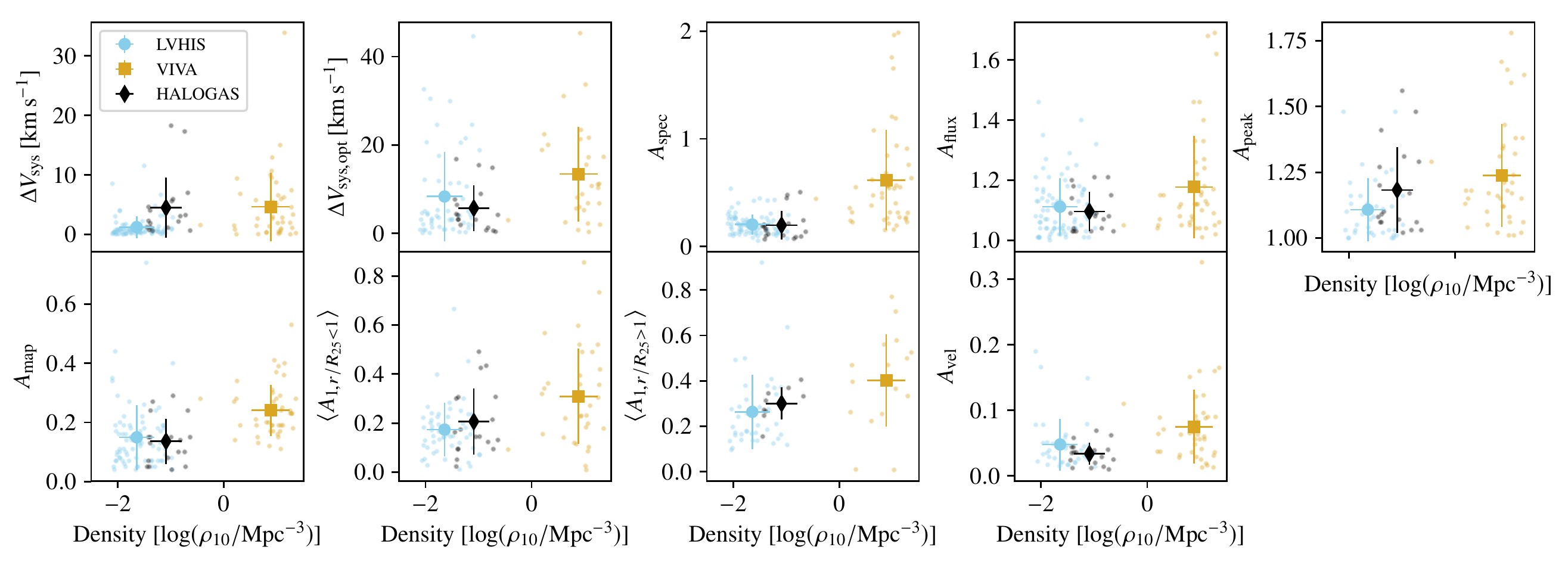}
		\caption{Asymmetry parameters vs environment density for LVHIS, VIVA and HALOGAS (blue circle, yellow square and black diamond, respectively). The large symbols indicate the mean of each sample with error bar indicating the standard deviation. From top left to bottom right: systemic velocity difference between flux weighted mean and $w_{50}/2$ ($\Delta V_{\mathrm{sys}}$), difference between H\,\textsc{i} and optical systemic velocities ($\Delta V_{\mathrm{sys,opt}}$), integrated spectrum residual ($A_{\mathrm{spec}}$), flux asymmetry ($A_{\mathrm{flux}}$), peak flux asymmetry ($A_{\mathrm{peak}}$), integrated intensity (moment 0) map residual ($A_{\mathrm{map}}$), average scaled first Fourier coefficient for the inner and outer disk ($\langle A_{1,r/R_{25}<1} \rangle$ and $\langle A_{1,r/R_{25}>1} \rangle$, respectively) and velocity field (moment 1) map weighted median absolute deviation scaled by the 95$^{\mathrm{th}}$ percentile ($A_{\mathrm{vel}}$).}
	\label{fig:env_asym_all}
\end{figure*}

Asymmetries are expected to be more common and stronger in denser environments as the frequency of galaxy interactions is higher and the IGM is likely to be denser, increasing the effectiveness of ram pressure stripping. The question is do we see this trend in the computed asymmetry parameters? In Fig.~\ref{fig:env_asym_all} we show the asymmetry parameters for LVHIS, VIVA and HALOGAS (blue circle, yellow square and black diamond, respectively) vs $10^{\mathrm{th}}$ nearest neighbour environment density for all galaxies and the mean and standard deviation for each survey (see Table~\ref{table:param_stats_all} for the mean and standard deviation values and Appendix~\ref{appendix:tabulated_params} for tabulated asymmetry parameters for all galaxies with the full tables available online). We focus on the influence of the environment density in this section before considering the effect of stellar mass in Section~\ref{s-sec:asym_mstar}.

The largest separation in asymmetry parameter between samples is $A_{\mathrm{spec}}$ (top centre panel in Fig.~\ref{fig:env_asym_all}) with the mean VIVA value $\sim3$ times larger than the LVHIS and HALOGAS values, which are similar. The VIVA galaxies cover a much larger range of measured $A_{\mathrm{spec}}$ values ($\sim$0.08--0.5) compared with LVHIS and HALOGAS ($\sim$0.00--0.16). Although we find a clear separation in the mean $A_{\mathrm{spec}}$ between low (LVHIS and HALOGAS) and high (VIVA) density environments, there are VIVA galaxies with small $A_{\mathrm{spec}}$ values overlapping the $A_{\mathrm{spec}}$ range seen for LVHIS and HALOGAS galaxies. Thus we cannot say the measured $A_{\mathrm{spec}}$ is completely a result of external mechanisms (i.e. galaxy-galaxy interactions and ram pressure stripping), but it is indicative that the environment is influencing the higher values measured.

Not showing as clear a separation as $A_{\mathrm{spec}}$, but still hinting at trends of larger asymmetry values at higher densities are $\Delta V_{\mathrm{sys}}$, $A_{\mathrm{map}}$, $A_{\mathrm{flux}}$, $\langle A_{1,r/R_{25}<1} \rangle$, $\langle A_{1,r/R_{25}>1} \rangle$ and $A_{\mathrm{vel}}$, while $\Delta V_{\mathrm{sys,opt}}$ and $A_{\mathrm{peak}}$ each cover similar ranges of asymmetry value at all densities. The large scatter and overlap between galaxy samples for these parameters demonstrates the wide distribution of galaxy asymmetries regardless of the environment density. However, we find the mean asymmetry tends to be higher in the denser environments suggesting external environmental mechanisms are influencing the measured asymmetry. The asymmetry parameters with possible trends with environment provide the best candidates to apply to future large surveys, such as WALLABY, which will have the larger galaxy samples required to be able either confirm or disprove these trends.

The lack of any trends in $\Delta V_{\mathrm{sys,opt}}$ and $A_{\mathrm{peak}}$ with density show that these parameters are not particularly meaningful for measuring environmental effects on asymmetry. There are also limitations in the ability to calculate these asymmetry parameters, which may be contributing to washing out any potential trends with environment. The H\,\textsc{i}--optical velocity difference, $\Delta V_{\mathrm{sys,opt}}$, requires optical redshift measurements, which do not always exist (e.g. for dwarf galaxies). Additionally, values sometimes differ between different surveys \citep[e.g. the optical redshifts for VIVA galaxies from SDSS do not all agree with the values taken from NED quoted in][]{Chung2009}. Ideally, calculating $\Delta V_{\mathrm{sys,opt}}$ requires a single optical redshift survey catalogue with measured redshifts corresponding to all H\,\textsc{i} detected galaxies. However, even using a single catalogue cannot remove intrinsic uncertainties in the measured redshifts, which are generally much larger than the measured $\Delta V_{\mathrm{sys,opt}}$ values \citep[e.g. $\Delta cz\sim46$\,km\,s$^{-1}$ for 6dFGS,][]{Jones2009}. The peak flux asymmetry ratio, $A_{\mathrm{peak}}$, can only be calculated for galaxies with clear double-horned spectral profiles. Hence, $A_{\mathrm{peak}}$ is useless for galaxies with a Gaussian shape and a single peak (i.e. many dwarf galaxies). 

\begin{table*}
	\centering
    \caption{Mean and standard deviation of asymmetry parameters for galaxies over the full stellar mass range ($6 \leq \log(M_*/\mathrm{M}_{\odot}) \leq 11$) and the subset of galaxies with stellar masses in the range $9 \leq \log(M_*/\mathrm{M}_{\odot}) \leq 10$. The there columns for each survey are the mean, $\langle x\rangle$, the $1\sigma$ standard deviation for the sample and the error on the mean, $\sigma/\sqrt{N}$, where $N$ is the number of galaxies.}
	\label{table:param_stats_all}
	\begin{tabular}{lccccccccr}
		\hline
		& \multicolumn{3}{c}{LVHIS} & \multicolumn{3}{c}{VIVA} & \multicolumn{3}{c}{HALOGAS} \\ \hline
		& $\langle x\rangle$ & $1\sigma$ & $\sigma/\sqrt{N}$ & $\langle x\rangle$ & $1\sigma$ & $\sigma/\sqrt{N}$ & $\langle x\rangle$ & $1\sigma$ & $\sigma/\sqrt{N}$ \\ \\
		& \multicolumn{9}{c}{$6 \leq \log(M_*/\mathrm{M}_{\odot}) \leq 11$} \\ \\
		$\Delta V_{\mathrm{sys}}$ [km\,s$^{-1}$]     & 1.2 & 1.9 & 0.2 & 4.6 & 5.8 & 0.8 & 4.5 & 5.1 & 1.0 \\
		$\Delta V_{\mathrm{sys,opt}}$ [km\,s$^{-1}$] & 8.4  & 10.1 & 1.1 & 13.4 & 10.7 & 1.6 & 5.7 & 5.2 & 1.0 \\
		$A_{\mathrm{spec}}$     & 0.204 & 0.095 & 0.011 & 0.615 & 0.465 & 0.068 & 0.197 & 0.134 & 0.027 \\
		$A_{\mathrm{flux}}$     & 1.11 & 0.10 & 0.01 & 1.18 & 0.17 & 0.02 & 1.10 & 0.07 & 0.01 \\
        $A_{\mathrm{peak}}$     & 1.11 & 0.12 & 0.01 & 1.24 & 0.20 & 0.03 & 1.18 & 0.16 & 0.03 \\
        $A_{\mathrm{map}}$      & 0.15 & 0.10 & 0.01 & 0.25 & 0.09 & 0.01 & 0.14 & 0.08 & 0.02 \\
        $\langle A_{1,r/R_{25}<1}\rangle$  & 0.17 & 0.11 & 0.01 & 0.31 & 0.19 & 0.03 & 0.21 & 0.14 & 0.03 \\
        $\langle A_{1,r/R_{25}>1}\rangle$  & 0.26 & 0.16 & 0.02 & 0.40 & 0.20 & 0.03 & 0.30 & 0.07 & 0.01 \\
        $A_{\mathrm{vel}}$    & 0.048 & 0.040 & 0.004 & 0.075 & 0.056 & 0.008 & 0.034 & 0.017 & 0.003 \\ \\
        & \multicolumn{9}{c}{$9 \leq \log(M_*/\mathrm{M}_{\odot}) \leq 10$} \\ \\
		$\Delta V_{\mathrm{sys}}$ [km\,s$^{-1}$]     & 1.2 & 1.2 & 0.3 & 2.2 & 2.3 & 0.6 & 2.8 & 2.6 & 1.3 \\
		$\Delta V_{\mathrm{sys,opt}}$ [km\,s$^{-1}$] & 7.6 & 8.4 & 2.8 & 15.9 & 11.4 & 2.9 & 1.9 & 1.4 & 0.7 \\
		$A_{\mathrm{spec}}$     & 0.216 & 0.106 & 0.028 & 0.795 & 0.618 & 0.150 & 0.145 & 0.059 & 0.030 \\
		$A_{\mathrm{flux}}$     & 1.11 & 0.07 & 0.02 & 1.19 & 0.17 & 0.04 & 1.07 & 0.05 & 0.02 \\
        $A_{\mathrm{peak}}$     & 1.10 & 0.09 & 0.03 & 1.20 & 0.21 & 0.06 & 1.11 & 0.11 & 0.05 \\
        $A_{\mathrm{map}}$      & 0.15 & 0.09 & 0.01 & 0.27 & 0.09 & 0.02 & 0.20 & 0.06 & 0.03 \\
        $\langle A_{1,r/R_{25}<1}\rangle$  & 0.21 & 0.11 & 0.04 & 0.31 & 0.23 & 0.07 & 0.18 & 0.07 & 0.04 \\
        $\langle A_{1,r/R_{25}>1}\rangle$  & 0.25 & 0.14 & 0.07 & 0.40 & 0.21 & 0.07 & 0.34 & 0.01 & 0.005 \\
        $A_{\mathrm{vel}}$     & 0.054 & 0.049 & 0.015 & 0.088 & 0.078 & 0.020 & 0.034 & 0.018 & 0.009 \\ \hline
	\end{tabular}
\end{table*}

\subsection{Asymmetry vs Stellar Mass}
\label{s-sec:asym_mstar}

\begin{figure*}
	\centering
	\includegraphics[width=17.7cm]{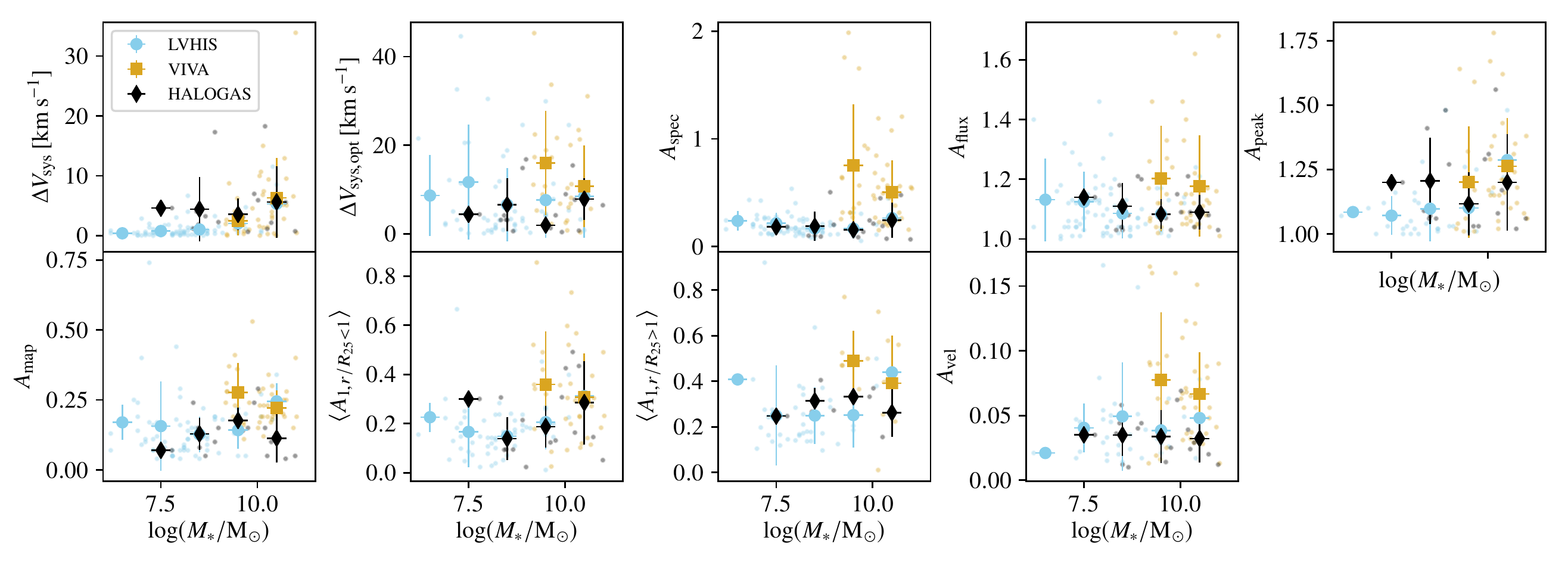}
		\caption{Same as Fig.~\ref{fig:env_asym_all}, but as a function of stellar mass instead of density.}
	\label{fig:mstar_asym}
\end{figure*}

Galaxy parameters should be compared among galaxies with similar properties (e.g. stellar mass as a proxy for halo mass). A galaxy's stellar mass will affect the impact of the environment on its symmetry. Higher stellar mass galaxies will have larger gravitational potentials and be more resistant to the influence of the environment on their matter distributions. Fig.~\ref{fig:mstar_asym} is the same as Fig.~\ref{fig:env_asym_all} except we show stellar mass on the $x$-axis rather than density and take the mean in 1\,dex stellar mass bins. We note that a drawback to binning by stellar mass is the decrease in the number of galaxies in each bin, which leads to low number statistics when taking the mean and standard deviation per bin. Larger galaxy samples (i.e. WALLABY) are required to better understand and disentangle the effect of stellar mass on these asymmetry parameters.

Similar to the $A_{\mathrm{spec}}$ vs density plot (top centre panel of Fig.~\ref{fig:env_asym_all}), the mean VIVA values are offset to larger values compared to the LVHIS and HALOGAS values in the same stellar mass bins (top centre panel of Fig.~\ref{fig:mstar_asym}). We do not find a stellar mass dependence on $A_{\mathrm{spec}}$ in the LVHIS or HALOGAS samples, neither of which vary with stellar mass. We next look at the other parameters which showed a tentative dependence on density ($\Delta V_{\mathrm{sys}}$, $A_{\mathrm{map}}$, $A_{\mathrm{flux}}$, $\langle A_{1,r/R_{25}<1} \rangle$, $\langle A_{1,r/R_{25}>1} \rangle$ and $A_{\mathrm{vel}}$) to try and disentangle any stellar mass effect on the measured asymmetry. We do not find any strong correlation between these parameters and stellar mass, especially compared to $A_{\mathrm{spec}}$. The variation with stellar mass of each sample and overlapping means at fixed stellar mass can be explained by the low numbers in each bin. This is particularly the case for the morphological and kinematic asymmetries which are not derived for all galaxies further reducing the numbers used for calculating the bin mean and standard deviation. Similar to environment density, the parameters $\Delta V_{\mathrm{sys,opt}}$ and $A_{\mathrm{peak}}$ show no trends with stellar mass. There may be a slight trend for higher stellar mass galaxies to have larger values for $\langle A_{1,r/R_{25}<1} \rangle$ and $\langle A_{1,r/R_{25}>1} \rangle$. However the small increase in the mean value is well within the standard deviation and could simply be due to low number statistics. 

\begin{figure}
	\centering
	\includegraphics[width=\columnwidth]{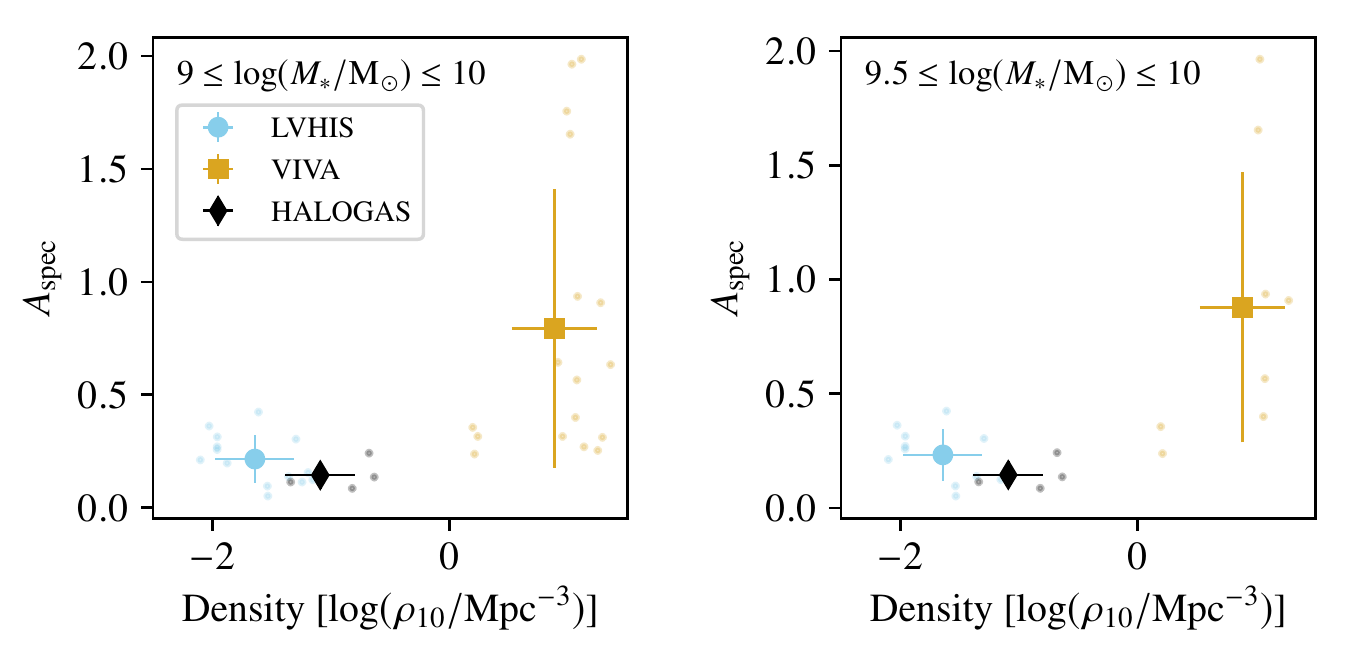}
		\caption{Spectrum residual asymmetry, $A_{\mathrm{spec}}$, vs environment density for LVHIS, VIVA and HALOGAS galaxies (blue circle, yellow square and black diamond, respectively) with stellar masses in the range $9 \leq \log(M_*/\mathrm{M}_{\odot}) \leq 10$ (left panel) and $9.5 \leq \log(M_*/\mathrm{M}_{\odot}) \leq 10$ (right panel). The large symbols indicate the mean of each sample with error bar indicating the standard deviation.}
	\label{fig:env_2param_m9}
\end{figure}

We also control for stellar mass by looking for trends in the mean asymmetry for each sample using galaxies in the stellar mass range $9 \leq \log(M_*/\mathrm{M}_{\odot}) \leq 10$. We plot $A_{\mathrm{spec}}$ vs density in Fig.~\ref{fig:env_2param_m9} (left and right panels for 1\,dex and 0.5\,dex stellar mass subsamples, respectively) and find no change within the $1\sigma$ standard deviations in the separation between the low (LVHIS and HALOGAS) and high (VIVA) densities. This supports the conclusion that the variation with density we see in the full samples (Fig.~\ref{fig:env_asym_all}) is due to density and not stellar mass. Similarly for the other asymmetry parameters, the trends in the mean for each galaxy 1\,dex stellar mass subsample remain roughly the same as for the full samples with any differences well within the standard deviation (see Table~\ref{table:param_stats_all} for the full and stellar mass selected samples, respectively). If we restrict the subsample to galaxies with stellar masses $9.5 \leq \log(M_*/\mathrm{M}_{\odot}) \leq 10$ we recover the same trends as the full sample and 1\,dex stellar mass subsample. This agrees with the lack of trends in asymmetry with stellar mass as we recover the same mean values when controlling for stellar mass. The absence of a stellar mass dependence in our results is in agreement with results from xGASS, which show no difference in the cumulative distribution of $A_{\mathrm{flux}}$ for galaxies with stellar masses above and below $\log(M_*/\mathrm{M}_{\odot})=10$ \citep{Watts2020}.

\subsection{Comparison with Published Results}
\label{s-sec:published_results}

As mentioned previously, many studies have investigated H\,\textsc{i} morphological and spectral asymmetries. In this section we compare our findings for various asymmetry parameters with results in the literature. 

\begin{figure}
	\centering
	\includegraphics[width=\columnwidth]{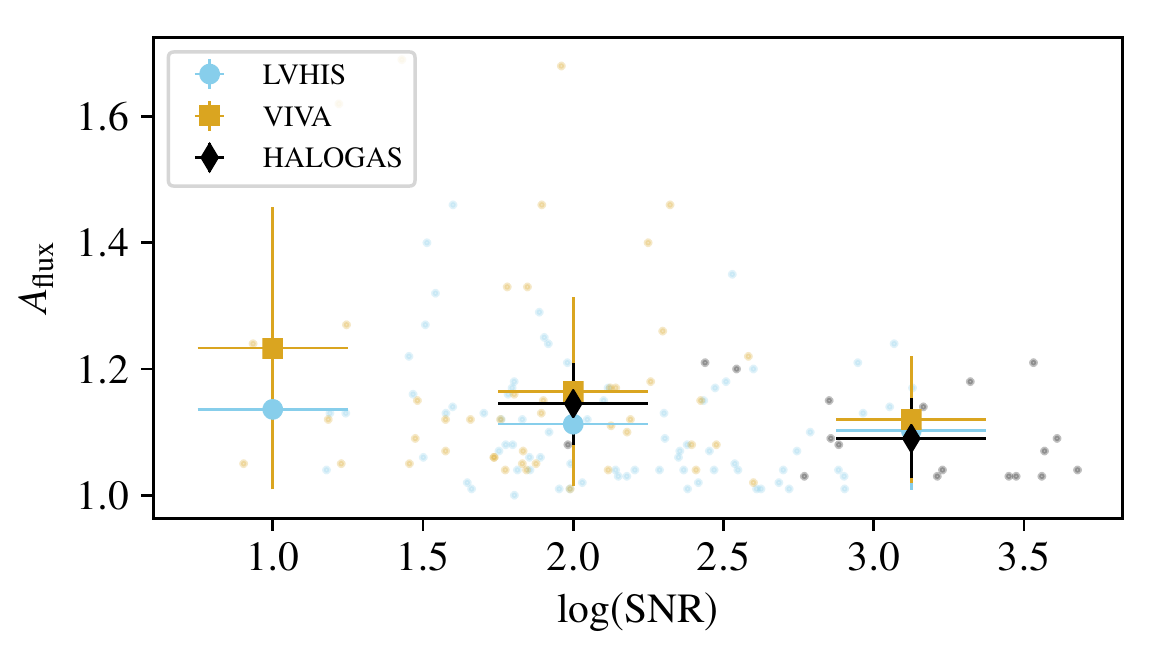}
		\caption{The flux ratio asymmetry, $A_{\mathrm{flux}}$, vs signal to noise ratio (SNR) for LVHIS, VIVA and HALOGAS galaxies (blue circle, yellow square and black diamond, respectively). The large symbols indicate the mean of each sample with error bar indicating the standard deviation.}
	\label{fig:snr_fasym}
\end{figure}

The flux ratio asymmetry, $A_{\mathrm{flux}}$, is the most commonly measured parameter as it can be computed for any galaxy with an integrated spectrum. If we define galaxies as asymmetric for $A_{\mathrm{flux}}>1.05$ (as used in early flux ratio asymmetry studies), we find asymmetric fractions of $38\%$, $44\%$ and $72\%$ for LVHIS, HALOGAS and VIVA, respectively, in agreement with the literature \citep[e.g. $45\%$, $77\%$ and $50\%$ in][respectively]{Richter1994,Matthews1998,Haynes1998}. However, of more interest is the fraction of galaxies with asymmetries potentially due to external mechanisms. For this we look for galaxies with $A_{\mathrm{flux}}>1.26$ and $>1.39$, corresponding to 2 and $3\sigma$ of a half Gaussian distribution determined by \cite{Espada2011} for isolated galaxies in the AMIGA project. The asymmetric fractions for LVHIS, VIVA and HALOGAS are $8\%$ ($3\%$), $20\%$ ($13\%$) and $0\%$ ($0\%$) of galaxies with $A_{\mathrm{flux}}>1.26$ ($>1.39$), corresponding to 6 (2), 9 (5) and 0 (0) galaxies, respectively. The group environment LVHIS and HALOGAS fractions are comparable to isolated, field galaxies \citep[$2\%$,][]{Espada2011} and much lower than for galaxy pairs \citep[$27\%$,][]{Bok2018}. This is unexpected as the LVHIS and HALOGAS samples probe galaxies in pairs and groups \citep[i.e.\ similar environments to those probed by][]{Bok2018}, not low density isolated, field environments of \cite{Espada2011}. However, we have not taken into account the differing signal to noise ratios (SNR) for the different studies. \cite{Watts2020} show that symmetric spectra will appear increasingly asymmetric for lower SNRs with the distribution in measured $A_{\mathrm{flux}}$ increasing by $\Delta A_{\mathrm{flux}}\sim0.05$--0.3 for $\mathrm{SNR}<50$. Our spectra are more robust to the effect of SNR as the majority of our spectra have $\mathrm{SNR}>50$ ($82\%$, $76\%$ and $100\%$ for LVHIS, VIVA and HALOGAS, respectively, Fig.~\ref{fig:optical_hists} lower, centre panel). We see no correlation between our measured asymmetry parameters and SNR, which we illustrate in Fig.~\ref{fig:snr_fasym} for $A_{\mathrm{flux}}$. We note the higher value of $A_{\mathrm{flux}}$ for HALOGAS in the middle SNR bin is a result of low number statistics, as this bin only contains two galaxies. An explanation for the lower asymmetric fractions we find is that the majority of the spectra used in the published results have $10<\mathrm{SNR}<50$, hence their $A_{\mathrm{flux}}$ are likely affected by the relatively lower SNRs and have a larger scatter compared with our values. Our trends of higher asymmetric fractions in denser environments is in agreement with the results from \cite{Watts2020} for xGASS satellites and centrals, with satellites found to be more frequently asymmetric than centrals and similarly for isolated vs group centrals.

At both the $2\sigma$ and $3\sigma$ cuts, the VIVA fractions of $20\%$ and $13\%$ are comparable to the Virgo and Abell\,1367 cluster results (16--$26\%$) of \cite{Scott2018}. Interestingly, \cite{Scott2018} measured $A_{\mathrm{flux}}$ in the Virgo cluster using integrated spectra from the AGES survey of 90 Virgo cluster galaxies with $\log(M_{\mathrm{HI}}/\mathrm{M}_{\odot})>8.48$, finding $16\%$ have $A_{\mathrm{flux}}>1.39$, and for 117 Virgo galaxies with $\log(M_{\mathrm{HI}}/\mathrm{M}_{\odot})\leq8.48$, finding $42\%$. The larger fractions found by \cite{Scott2018} at lower H\,\textsc{i} mass indicate that there is a mass dependence on $A_{\mathrm{flux}}$. However, similar to stellar mass, we find no significant variation in asymmetry with H\,\textsc{i} mass, which may be due to the smaller sample size used in this work (i.e.\ 136 galaxies).

\begin{figure*}
	\centering
	\includegraphics[width=17.7cm]{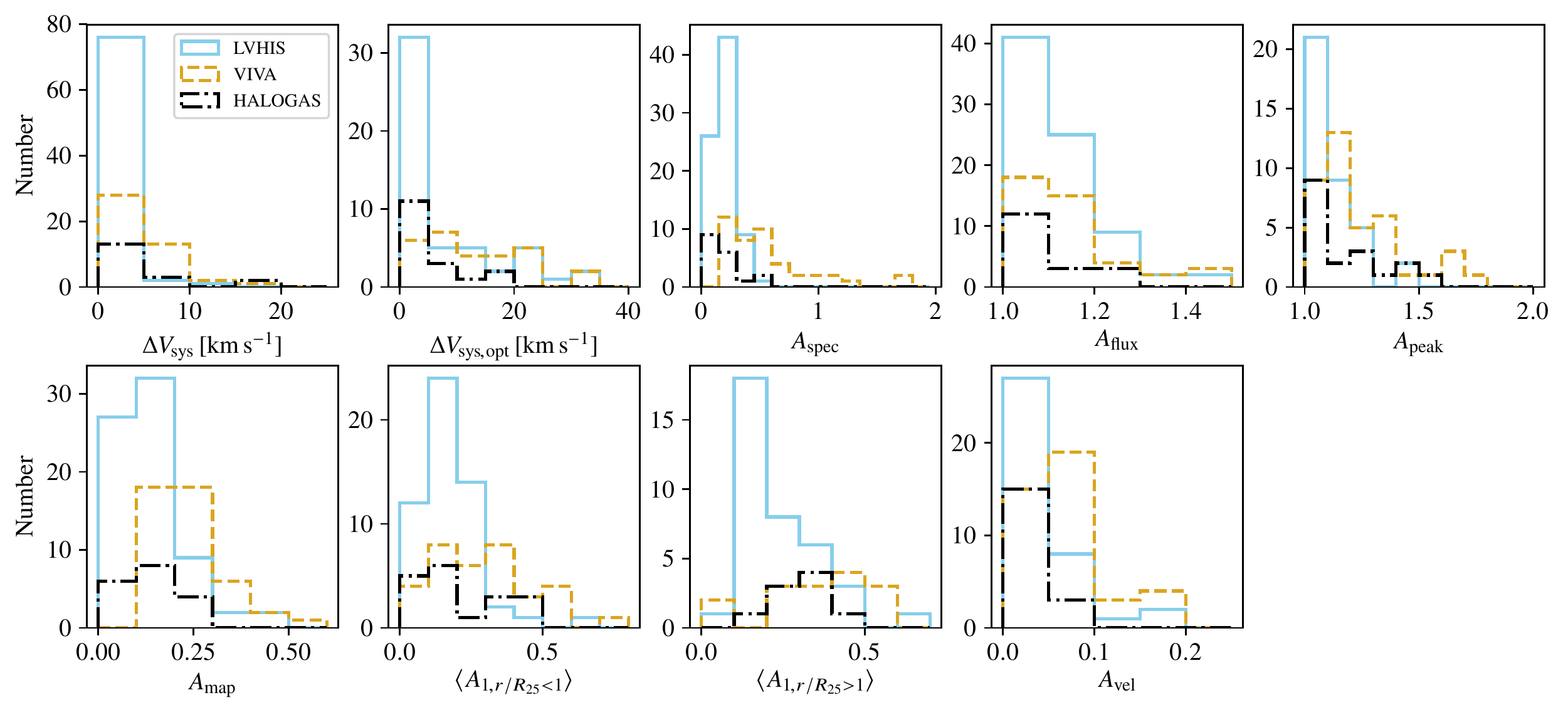}
		\caption{Similar to Fig.~\ref{fig:env_asym_all}, but showing histograms of asymmetry parameters.}
	\label{fig:asym_hist}
\end{figure*}

Fourier analysis has been applied to both near-IR and H\,\textsc{i} images, however these results cannot be directly compared as the asymmetry parameter, $\langle A_1 \rangle$, probes different locations within the studied galaxies. In near-IR analysis, $\langle A_1 \rangle$ is measured over 1--$2.5R_{\mathrm{scale}}$, where $R_{\mathrm{scale}}$ is the galaxy scale length \citep[e.g.][]{Bournaud2005b,Angiras2006,Angiras2007}. In H\,\textsc{i} analysis, $\langle A_1 \rangle$ can be measured out to the edge of the detected H\,\textsc{i} disk (i.e. to the column density sensitivity limit), which is generally a few times the optical radius, $R_{25}$. Hence in H\,\textsc{i}, $\langle A_1 \rangle$ can be measured to much larger radii compared to the near-IR, as $R_{25}$ is generally 4--5 times $R_{\mathrm{scale}}$ \citep{vanderKruit1982}. The mean value for $\langle A_{1,r/R_{25}<1} \rangle$ (inner disk) are similar or higher (0.17, 0.32 and 0.24 for LVHIS, VIVA and HALOGAS, respectively) than for isolated and group galaxies measured over 1--$2.5R_{\mathrm{scale}}$ \citep[0.11, 0.24 and 0.14 from][respectively]{Bournaud2005b,Angiras2006,Angiras2007}. The similar or small increases in the LVHIS and HALOGAS samples compared with the near-IR values can be attributed to measuring over a larger area of the galaxy disks as $A_1$ generally increases with radius \citep[e.g.][]{Angiras2006,Angiras2007,vanEymeren2011b}. The significantly higher mean for the VIVA sample is more likely to be an actual effect of the increased density in the cluster environment. Our results are directly comparable to the Fourier analysis of WHISP galaxies by \cite{vanEymeren2011b}, as we adopt their measured $\langle A_1 \rangle$ regimes: $r/R_{25}<1$ and $r/R_{25}>1$, inner and outer disk, respectively. The mean values of LVHIS, VIVA and HALOGAS for both the inner (0.17, 0.32 and 0.24) and outer (0.25, 0.40, 0.33) disks are larger than the WHISP values (0.11 and 0.15, respectively). In agreement with the WHISP values, the mean for the outer disk is larger than the inner disk for all galaxy samples (Fig.~\ref{fig:asym_hist}, lower left-centre and lower centre panels). \cite{vanEymeren2011b} classified many of the WHISP galaxies as isolated based on their criteria for classifying environment. Assuming the WHISP galaxies are isolated, we have probed higher density environments with LVHIS, VIVA and HALOGAS, which points to density as the possible driver of the higher measured $\langle A_1 \rangle$ values.

\subsection{Comparison of Asymmetry Measures}
\label{s-sec:asym_comp}

In addition to seeing how asymmetry parameters relate to environment and galaxy properties, it is useful to investigate and understand the correlations among the parameters themselves as they probe different types of asymmetry. The most useful correlations would be ones to relate parameters from integrated spectra to morphological and kinematic parameters as the majority of galaxies detected by upcoming surveys will be spatially unresolved, limiting the measurable parameters to the integrated spectrum. Correlations between spatially resolved and unresolved asymmetry parameters would provide the ability to relate nearby, spatially resolved galaxies to more distant, spatially unresolved galaxies. 

\begin{figure*}
	\centering
	\includegraphics[width=17.5cm]{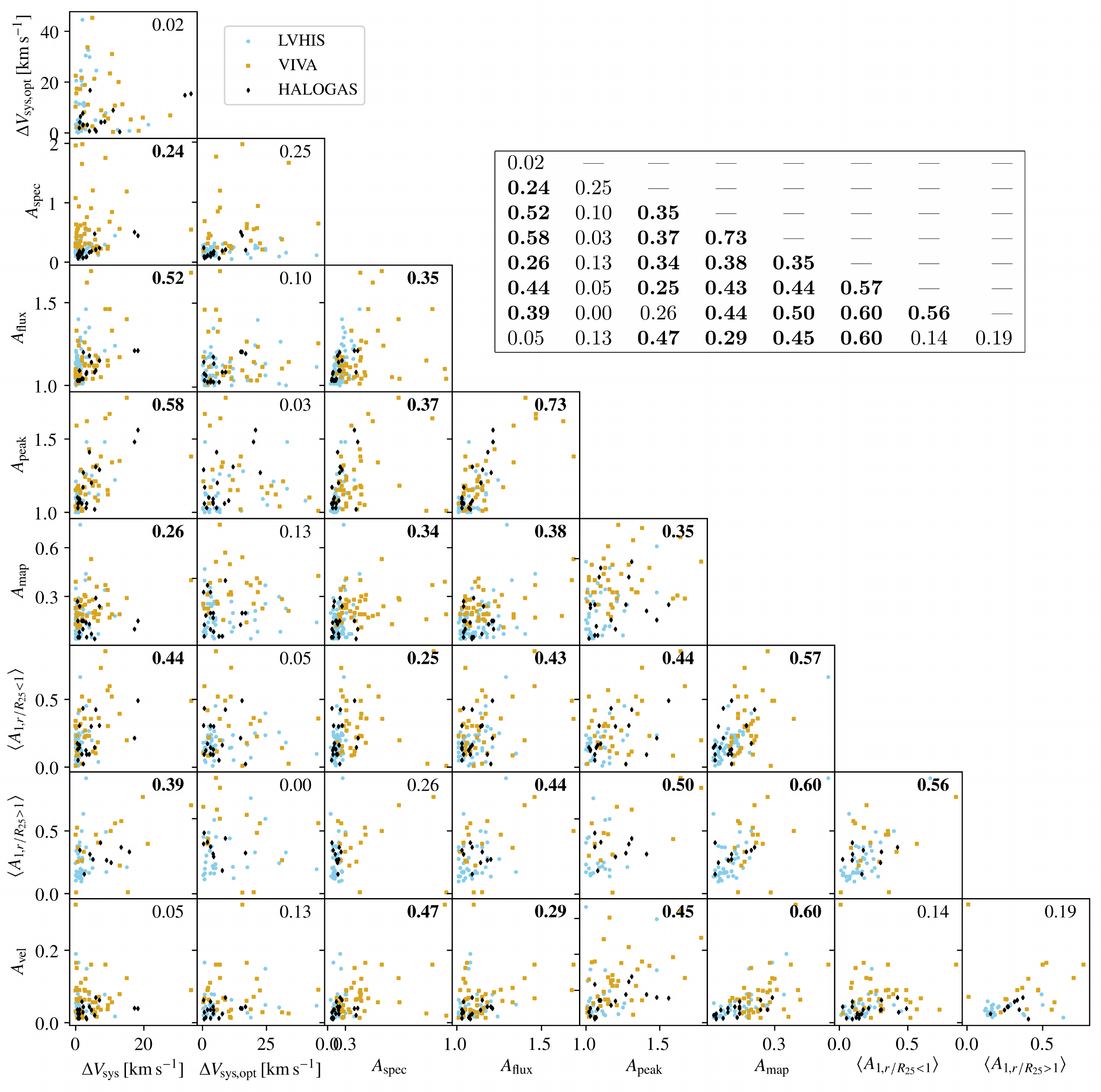}
		\caption{Correlation between each pair of asymmetry parameters for the combined LVHIS, VIVA and HALOGAS data sets with the Pearson correlation coefficients shown in top right of each panel (also shown in the inset table). We have highlighted significant Pearson correlations (2-tailed p-value $<0.01$) in bold.}
	\label{fig:param_corr}
\end{figure*}

We calculate the Pearson correlations for the full combined sample of the LVHIS, VIVA and HALOGAS data sets. We plot each parameter pair and list the correlation coefficient in the top right corner of each panel in Fig.~\ref{fig:param_corr} with significant Pearson correlations (2-tailed p-value $<0.01$) highlighted in bold. Unsurprisingly, the strongest correlations are between parameters directly related to each other. The integrated spectrum asymmetry parameters of $\Delta V_{\mathrm{sys}}$, $A_{\mathrm{spec}}$, $A_{\mathrm{flux}}$ and $A_{\mathrm{peak}}$ have moderate correlations $>0.23$, while the morphological parameters of $A_{\mathrm{map}}$, $\langle A_{1,r/R_{25}<1} \rangle$ and $\langle A_{1,r/R_{25}>1} \rangle$ have stronger correlations $>0.5$. We find moderate correlations between integrated spectrum and morphological asymmetries. $A_{\mathrm{spec}}$ has moderate correlations of $>0.24$ with $A_{\mathrm{map}}$ and $\langle A_{1,r/R_{25}<1} \rangle$. $A_{\mathrm{flux}}$ and $\Delta V_{\mathrm{sys}}$ moderately correlate with $A_{\mathrm{map}}$, $\langle A_{1,r/R_{25}<1} \rangle$ ($>0.34$) and $\langle A_{1,r/R_{25}>1} \rangle$. The kinematic asymmetry, $A_{\mathrm{vel}}$, has statistically significant correlations with $A_{\mathrm{spec}}$, $A_{\mathrm{map}}$ and $A_{\mathrm{peak}}$. These correlations confirm that asymmetries in the H\,\textsc{i} kinematics and distribution within a galaxy affect the measured asymmetries in the galaxy's integrated spectrum, which can provide an indication of the H\,\textsc{i} distribution in spatially unresolved galaxies. The weaker correlation of $A_{\mathrm{flux}}$ and $\Delta V_{\mathrm{sys}}$ with $\langle A_{1,r/R_{25}>1} \rangle$ indicates that these spectral asymmetry measures are less sensitive to asymmetries at larger radii, where the H\,\textsc{i} is more diffuse, as the spectrum is dominated by the denser, larger mass contribution at smaller radii.

\subsection{Origin of Asymmetries}
\label{s-sec:origin}

As mentioned in Section~\ref{sec:intro}, there are a number of proposed mechanisms for producing observed asymmetries. In low to high density group and cluster environments, tidal interactions \citep{Jog1997}, galaxy mergers \citep[e.g.][]{Bok2018} and ram pressure stripping \citep[e.g.][]{Mapelli2008} are the most likely causes of H\,\textsc{i} asymmetries. Although ram pressure is more effective in galaxy clusters due to the higher density of the intergalactic medium (IGM) and higher relative galaxy velocities, it is also observed in lower density environments \citep[e.g.][]{Rasmussen2006,Westmeier2011,Rasmussen2012}. Isolated galaxies, without any nearby neighbours, also display asymmetries, which are proposed to be caused by gas accretion along filaments \citep[e.g.][]{Bournaud2005b,Mapelli2008,LagosP2018}, minor mergers of satellites \citep[e.g.][]{Zaritsky1997,Bournaud2005b} and fly-by interactions \citep[e.g.][]{Mapelli2008}.

In addition to external mechanisms, there are also internal mechanisms proposed to lead to asymmetries \citep[e.g. non-circular motions and $m=1$ spiral waves,][respectively]{Baldwin1980,Masset1997}. However, it is unlikely that internal mechanisms are giving rise to the asymmetries we measure in this work. Most of the LVHIS, VIVA and HALOGAS galaxies have $\langle A_{1,r/R_{25}<1} \rangle$ and $\langle A_{1,r/R_{25}>1} \rangle$ values $>0.05$ implying the observed asymmetries are external in origin as modelling finds that internal mechanisms only produce $\langle A_{1,r/R_{25}<1} \rangle$ and $\langle A_{1,r/R_{25}>1} \rangle$ values $<0.05$ \citep{Bournaud2005b}.

The slight trends of larger mean asymmetries we find in galaxies in higher density environments suggest that tidal interactions and ram pressure stripping are likely to be the main causes of the observed asymmetries. Major mergers are unlikely to be responsible for the asymmetries observed, as the LVHIS, VIVA and HALOGAS samples contain late-type, irregular and dwarf galaxies, not early-type galaxies which are thought to form due to major mergers \citep[e.g.][]{Toomre1977}. Simulations show that major mergers lead to galaxies with earlier Hubble types and exhibit higher measured asymmetries during the merging process, while post-merger the final galaxy becomes more symmetric with lower measured asymmetries after $\sim1$\,Gyr \citep{Walker1996,Bournaud2005a,Bournaud2005b}. However, minor mergers are still a possible cause. Gravitational interactions and minor mergers produce asymmetries with longer life-times of $\sim2$--4\,Gyr in simulations \citep{Bournaud2005b}, which can provide a time frame for a galaxy's interaction history based on measured asymmetries. 

Ram pressure likely has the greatest effect on the VIVA asymmetries as this sample probes the cluster environment of Virgo. However, several instances of likely tidal interactions causing asymmetries in VIVA galaxies have been identified \citep{Chung2009}. Asymmetries in LVHIS and HALOGAS are likely predominantly caused by tidal interactions and possibly mergers as these samples probe pair and group environments. However, \cite{deblok2014} and \cite{Westmeier2011} have also identified ram pressure stripping as affecting galaxies in HALOGAS (NGC\,4414) and LVHIS (NGC\,300), respectively. Ram pressure would be more likely to affect HALOGAS and LVHIS galaxies in larger groups (e.g. Sculptor and Coma I) with a denser IGM, more similar to in clusters, compared to the low density IGM around the galaxy pairs probed by these samples. The LVHIS, VIVA and HALOGAS galaxy samples do not probe isolated galaxies, so gas accretion probably has a small influence on the measured asymmetries as there is likely less cold gas in higher density environments \citep[e.g.][]{Angiras2006}. However, inflows could be responsible for the asymmetries measured in the lowest density LVHIS galaxies, as these galaxies are unlikely to be undergoing tidal interactions and the IGM is likely to be very low in density.

\subsection{Implications for WALLABY}
\label{s-sec:wallaby_implications}

WALLABY will detect $\sim500\,000$ galaxies in H\,\textsc{i} out to $z\sim0.26$ across $\sim75\%$ of the sky, enabling WALLABY to probe environment densities ranging from isolated, field galaxies to dense cluster environments in statistically significant numbers. Compared with the analysis here, with WALLABY we will have sufficient numbers of detections to more finely bin galaxies by density while also spanning a much larger range in environment densities. 

\begin{figure}
	\centering
	\includegraphics[width=\columnwidth]{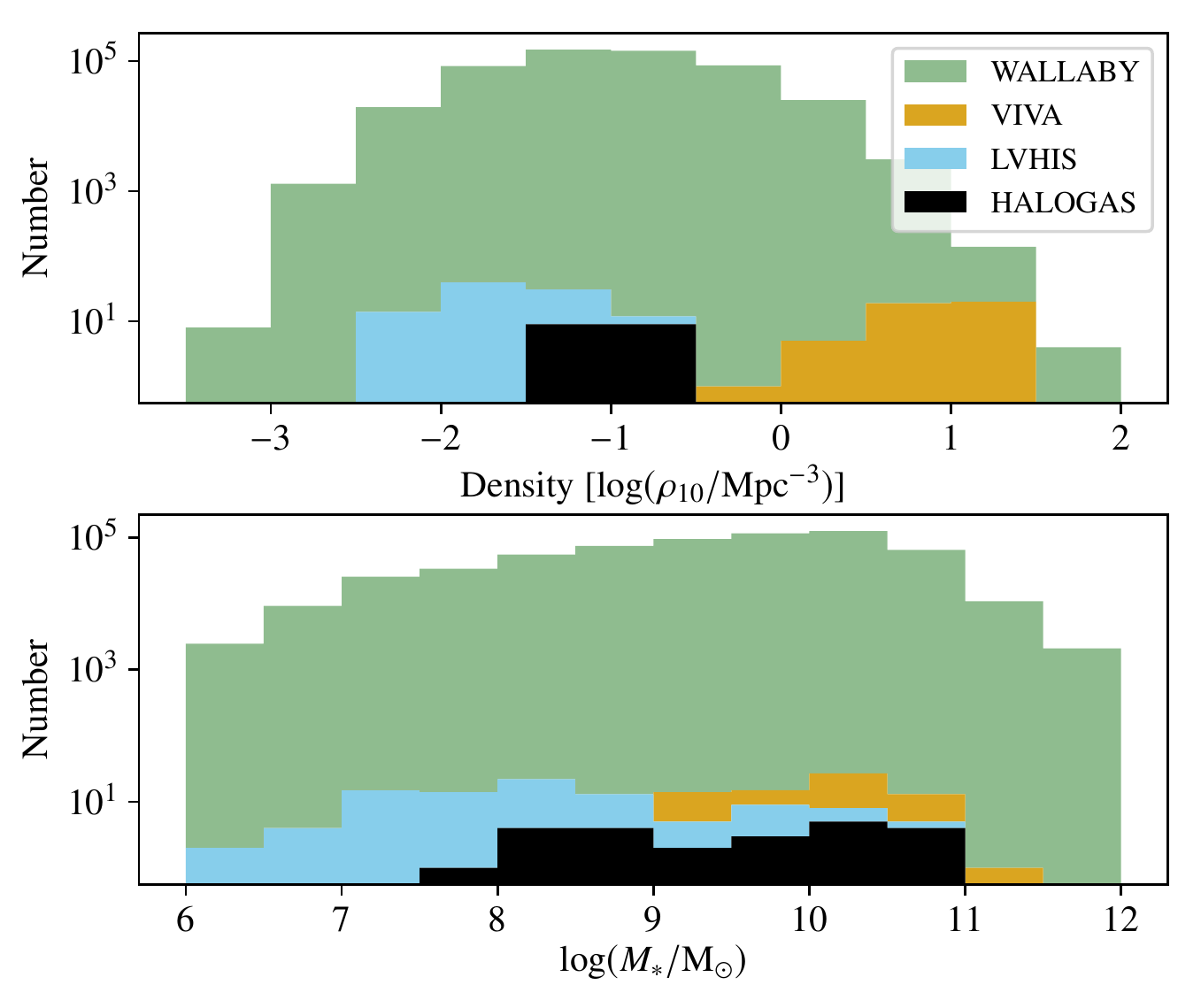}
		\caption{Histogram of $10^{\mathrm{th}}$ nearest neighbour environment density and stellar mass (upper and lower panels, respectively) for mock WALLABY detections and LVHIS, VIVA and HALOGAS galaxies (green, blue, orange and black, respectively).}
	\label{fig:wallaby_hists}
\end{figure}

To provide predictions for WALLABY, we create a mock survey catalogue using the lightcone created from the medi-SURFS simulation box described in Section~\ref{s-sec:environment}. We also create a lightcone from a smaller SURFS box, denoted micro-SURFS (40\,cMpc/h on a side and $512^3$ dark matter particles), using the same method as previously discussed. Micro-SURFS has a higher resolution than medi-SURFS (dark matter particle mass resolutions: $4.13\times10^7\,\mathrm{M}_{\odot}$/h and $2.21\times10^8\,\mathrm{M}_{\odot}$/h for micro- and medi-SURFS, respectively). To get the final mock survey catalogue we combine the micro-SURFS detections for $z<0.04$ and $M_*>10^6\,\mathrm{M}_{\odot}$ with the medi-SURFS detections for $z>0.04$ and $M_*>10^8\,\mathrm{M}_{\odot}$, as micro-SURFS resolves smaller galaxies which will only be detectable with WALLABY in the nearby Universe (the applied mass limits correspond to the resolution limits of each SURFS box). We then create H\,\textsc{i} emission for each mock galaxy using the H\,\textsc{i} emission line generator code\footnote{\url{https://github.com/garimachauhan92/HI-Emission-Line-Generator}} presented in \cite{Chauhan2019}. We estimate the WALLABY detections by calculating the integrated flux of each mock galaxy, using the atomic mass estimated as an output of \textsc{shark} and the distance from the lightcone, and comparing it with the sensitivity of ASKAP \citep{Duffy2012}. We also find that the integrated flux sensitivity is heavily reliant on the $w_{50}$ of the emission lines, which is calculated by the emission line code. For each mock WALLABY detection in medi-SURFS we compute the weighted 10$^{\mathrm{th}}$ nearest neighbour density using the full medi-SURFS mock catalogue to produce consistent environment densities comparable to the corrected LVHIS, VIVA and HALOGAS densities (see Section~\ref{s-sec:environment}). We do not calculate densities for mock WALLABY detections in micro-SURFS (i.e. galaxies with $M_*<10^8\,\mathrm{M}_{\odot}$) as there is no relation between galaxy positions in micro-SURFS and medi-SURFS, which would produce incorrect micro-SURFS galaxy densities. Fig.~\ref{fig:wallaby_hists} shows histograms of the density and stellar mass distributions for the mock WALLABY detections. WALLABY will detect $\sim10^3$--$10^5$ galaxies in 0.5\,dex bins over $\sim$5 orders of magnitude in both environment density and stellar mass.

Compared with the LVHIS, VIVA and HALOGAS samples, WALLABY will detect 1--3 orders of magnitude more spatially unresolved and $\sim1$ order of magnitude more spatially resolved galaxies at similar environment densities, and similarly in stellar mass. This will provide the ability to reduce the dispersion in asymmetry at fixed density and stellar mass and strengthen or disprove the tentative trends we observe and can be used to associate observed asymmetries with internal vs external mechanisms. Similarly, WALLABY will provide much more uniform coverage in stellar mass than in this work and will be able to disentangle the effect of stellar mass from environment using more refined bins, containing more galaxies, than is possible in this work.

Spectral asymmetries can be derived for all WALLABY detections, while morphological and kinematic asymmetries can only be calculated for those WALLABY detections spatially resolved by $\geq3$ beams. Koribalski et al. (in prep.) predict $\sim5\,000$ WALLABY detections will be resolved by $\geq5$ WALLABY beams, providing a lower limit on the number of galaxies with spatial resolutions $\geq3$ beams, however the physical scales probed in each detection will not be the same. Probing morphological and kinematic asymmetries requires observations with similar physical scales, as decreasing resolution can smooth out flux and velocities or produce offsets in the H\,\textsc{i} defined centre of the galaxy \citep[e.g.][]{Scott2018}. Nearby galaxies with higher physical scale resolution can be smoothed to match the lower resolutions of more distant galaxies, which will increase the sample size for direct comparison of morphological and kinematic asymmetries and enable the comparison of galaxies at a greater range of distances. Even with this limitation, WALLABY will still provide significantly larger samples covering a range of resolved physical scales than past surveys.

\section{SUMMARY}
\label{sec:conclusion}

In this work we have investigated the influence of environment density and stellar mass on measured spectral asymmetry parameters from integrated spectra and morphological and kinematic asymmetry parameters from spatially resolved images of galaxies in the LVHIS, VIVA and HALOGAS surveys. Our main results are:
\begin{itemize}
    \item We find a trend in the integrated spectrum residual with environment density and a hint of trends in the morphological asymmetry parameters, the weighted median absolute deviation of the velocity field, the flux ratio asymmetry and the difference in measured H\,\textsc{i} systemic velocities with environment. The environmental dependence is also supported from comparison of morphological asymmetries presented here with previously published results. However, larger galaxy samples are required to determine if these are true trends or artefacts of low number statistics.
    
    \item We find no correlations between H\,\textsc{i} asymmetries and either stellar or H\,\textsc{i} mass. The lack of correlation with H\,\textsc{i} mass is in disagreement with the results of \cite{Scott2018} for Virgo cluster galaxies. \cite{Scott2018} found a larger asymmetric fraction (42\%) of galaxies with H\,\textsc{i} masses $\log(M_{\mathrm{HI}}/\mathrm{M}_{\odot})\leq8.48$ compared with 16\% with $\log(M_{\mathrm{HI}}/\mathrm{M}_{\odot})>8.48$ (117 and 90 galaxies, respectively) and is potentially due to the small sample size used in this work (136 galaxies).
    
    \item As expected, the strongest correlations are between related asymmetry measures (i.e. between spectral parameters). However, we also find moderate correlations of spectral asymmetries with morphological and kinematic asymmetry parameters indicating spatially resolved asymmetries influence the asymmetries measured from integrated spectra.
    
    \item We propose parameterising integrated spectrum asymmetries with the spectrum residual ($A_{\mathrm{spec}}$), H\,\textsc{i} systemic velocity difference ($\Delta V_{\mathrm{sys}}$) and flux ratio asymmetry ($A_{\mathrm{flux}}$). For spatially resolved galaxies, we propose parameterising morphological asymmetries with the integrated intensity map residual ($A_{\mathrm{map}}$) and Fourier analysis of the integrated intensity map ($\langle A_{1,r/R_{25}<1} \rangle$ and $\langle A_{1,r/R_{25}>1} \rangle$) and kinematic asymmetries using the weighted median absolute deviation of the velocity field scaled by the 95$^{\mathrm{th}}$ percentile ($A_{\mathrm{vel}}$).
    
    \item Tidal interactions and ram pressure stripping are the most likely mechanisms responsible for the measured asymmetries, while gas inflows could cause the small measured asymmetries in the LVHIS galaxies at the lowest estimated environment densities. This is supported by LVHIS, VIVA and HALOGAS galaxies identified as undergoing tidal interactions and ram pressure stripping in previous work. However, the statistical analysis we present here cannot be used to distinguish between tidal interactions and ram pressure, which requires detailed, targeted studies of individual spatially resolved systems.
\end{itemize}

Upcoming surveys, such as WALLABY, will detect significantly larger numbers of galaxies covering a wide range of environment densities and stellar masses. These surveys will provide single uniform samples for which we can carry out the analysis presented here. This will remove inherent differences between surveys, be matched to a single optical redshift catalogue for environment estimation and remove the influence of low number statistics. Future studies will then be able to investigate correlations between H\,\textsc{i} asymmetries with other galaxy properties in addition to environment and stellar mass, such as H\,\textsc{i} deficiency and star formation rate.

\section*{Acknowledgements}

We thank B.~Koribalski for providing helpful comments. We also thank B.~Holwerda for instructive discussions and comments. This research was conducted by the Australian Research Council Centre of Excellence for All Sky Astrophysics in 3 Dimensions (ASTRO 3D), through project number CE170100013. This research was supported by the Munich Institute for Astro- and Particle Physics (MIAPP) of the DFG cluster of excellence "Origin and Structure of the Universe This research has made use of the NASA/IPAC Extragalactic Database (NED) which is operated by the Jet Propulsion Laboratory, California Institute of Technology, under contract with the National Aeronautics and Space Administration. The Local Volume H\,\textsc{i} Survey H\,\textsc{i} 21-cm observations were obtained with the Australia Telescope Compact Array which is part of the Australia Telescope National Facility that is funded by the Commonwealth of Australia for operations as a National Facility managed by Commonwealth Scientific and Industrial Research Organisation (CSIRO). This research made use of data from WSRT HALOGAS-DR1. The Westerbork Synthesis Radio Telescope is operated by ASTRON (Netherlands Institute for Radio Astronomy) with support from the Netherlands Foundation for Scientific Research NWO. This research used data from VIVA survey carried out on the Karl G. Jansky Very Large Array, which is part of the National Radio Astronomy Observatory. The National Radio Astronomy Observatory is a facility of the National Science Foundation operated under cooperative agreement by Associated Universities, Inc.

Funding for the Sloan Digital Sky Survey IV has been provided by the Alfred P. Sloan Foundation, the U.S. Department of Energy Office of Science, and the Participating Institutions. SDSS-IV acknowledges support and resources from the Center for High-Performance Computing at the University of Utah. The SDSS web site is www.sdss.org.

SDSS-IV is managed by the Astrophysical Research Consortium for the Participating Institutions of the SDSS Collaboration including the Brazilian Participation Group, the Carnegie Institution for Science, Carnegie Mellon University, the Chilean Participation Group, the French Participation Group, Harvard-Smithsonian Center for Astrophysics, Instituto de Astrof\'isica de Canarias, The Johns Hopkins University, Kavli Institute for the Physics and Mathematics of the Universe (IPMU) / University of Tokyo, the Korean Participation Group, Lawrence Berkeley National Laboratory, Leibniz Institut f\"ur Astrophysik Potsdam (AIP), Max-Planck-Institut f\"ur Astronomie (MPIA Heidelberg), Max-Planck-Institut f\"ur Astrophysik (MPA Garching), Max-Planck-Institut f\"ur Extraterrestrische Physik (MPE), National Astronomical Observatories of China, New Mexico State University, New York University, University of Notre Dame, Observat\'ario Nacional / MCTI, The Ohio State University, Pennsylvania State University, Shanghai Astronomical Observatory, United Kingdom Participation Group, Universidad Nacional Aut\'onoma de M\'exico, University of Arizona, University of Colorado Boulder, University of Oxford, University of Portsmouth, University of Utah, University of Virginia, University of Washington, University of Wisconsin, 
Vanderbilt University, and Yale University.




\bibliographystyle{mnras}
\bibliography{master}



\appendix

\section{TABULATED ASYMMETRY PARAMETERS}
\label{appendix:tabulated_params}

Description of columns in Tables~\ref{table:param_lvhis}, \ref{table:param_viva} and \ref{table:param_halogas}.

\renewcommand{\labelenumi}{\arabic{enumi}}
\begin{enumerate}
    \item Galaxy ID
    \item $\log(\rho_{10}/\mathrm{Mpc}^{-3})$ -- environment density computed using the 10 nearest neighbours (Section~\ref{s-sec:environment})
    \item $\log\left(M_*/\mathrm{M}_{\mathrm{\odot}}\right)$ -- stellar mass
    \item $\Delta V_{\mathrm{sys}}$ -- difference between the flux weighted mean systemic velocity and the systemic velocity defined as the midpoint of the spectrum at $50\%$ of the spectrum's peak height (Section~\ref{ss-sec:vel_diff})
    \item $\Delta V_{\mathrm{sys,opt}}$ -- difference between the flux weighted mean systemic velocity from the H\,\textsc{i} spectrum and the optical velocity (Section~\ref{ss-sec:vel_diff})
    \item $A_{\mathrm{spec}}$ -- spectrum residual, the sum of the absolute differences between the flux in each channel of the spectrum and the flux in the spectral channel of the spectrum flipped about the systemic velocity normalised by the integrated flux of the spectrum (Section~\ref{ss-sec:spec_res})
    \item $A_{\mathrm{flux}}$ -- ratio of the integrated flux in the left and right halves of the spectrum divided at the systemic velocity (Section~\ref{ss-sec:flux_asym})
    \item $A_{\mathrm{peak}}$ -- ratio between the left and right peaks of the spectrum (Section~\ref{ss-sec:peak_asym})
    \item $A_{\mathrm{map}}$ -- bias corrected asymmetry parameter of integrated intensity (moment 0) map (Section~\ref{ss-sec:mom0_asym})
    \item $\langle A_{1}\rangle$ -- mean first harmonic coefficient from Fourier analysis of moment 0 map over radii smaller than the optical radius, $\left(\frac{r}{R_{25}}<1\right)$ (Section~\ref{ss-sec:fourier})
    \item $\langle A_{2}\rangle$ -- mean first harmonic coefficient from Fourier analysis of moment 0 map over radii larger than the optical radius, $\left(\frac{r}{R_{25}}>1\right)$ (Section~\ref{ss-sec:fourier})
    \item $A_{\mathrm{vel}}$ -- weighted median absolute deviation of the sum of the velocity field and the velocity field rotated by 180 degrees about the galaxy centre scaled by the 95th percentile of the difference map between the original and rotated velocity fields (Section~\ref{ss-sec:vel_mad})
\end{enumerate}

\begin{table*}
	\centering
    \caption{Asymmetry parameters for LVHIS galaxies. The full table is available online.}
	\label{table:param_lvhis}
	\begin{tabular}{lccccccccccr}
	    \\ \hline
        Galaxy & $\log(\rho_{10}/$ & $\log\left(\frac{M_*}{\mathrm{M}_{\mathrm{\odot}}}\right)$ & $\Delta V_{\mathrm{sys}}$ & $\Delta V_{\mathrm{sys,opt}}$ & $A_{\mathrm{spec}}$ & $A_{\mathrm{flux}}$ & $A_{\mathrm{peak}}$ & $A_{\mathrm{map}}$ & $\langle A_{1}\rangle$ & $\langle A_{2}\rangle$ & $A_{\mathrm{vel}}$ \\
        LVHIS & Mpc$^{-3}$) & & [km\,s$^{-1}$] & [km\,s$^{-1}$] & & & & & $\left(\frac{r}{R_{25}}<1\right)$ & $\left(\frac{r}{R_{25}}>1\right)$ & \\ \hline
        001 & $-1.37$ & 7.3 & 0 & 11 & 0.264 & 1.06 & --- & 0.07 & --- & --- & --- \\ 
        003 & $-1.24$ & 6.2 & 1 & 22 & --- & --- & --- & --- & --- &    --- & --- \\ 
        004 & $-1.19$ & 9.4 & 3 & 5 & --- & --- & 1.26 & 0.13 & 0.45 & --- & 0.032 \\ 
        005 & $-1.15$ & 9.5 & 0 & 4 & 0.124 & 1.03 & 1.01 & 0.19 & 0.15 & --- & 0.028 \\ 
        006 & $-1.50$ & 10.4 & 14 & 3 & 0.446 & 1.24 & 1.22 & 0.18 & --- & --- & 0.017 \\ 
        ... & ... & ... & ... & ... & ... & ... & ... & ... & ... & ... & ... \\ \hline
	\end{tabular}
\end{table*}

\begin{table*}
	\centering
    \caption{Asymmetry parameters for VIVA galaxies. The full table is available online.}
	\label{table:param_viva}
	\begin{tabular}{lccccccccccr}
	    \\ \hline
        Galaxy & $\log(\rho_{10}/$ & $\log\left(\frac{M_*}{\mathrm{M}_{\mathrm{\odot}}}\right)$ & $\Delta V_{\mathrm{sys}}$\, & $\Delta V_{\mathrm{sys,opt}}$\, & $A_{\mathrm{spec}}$ & $A_{\mathrm{flux}}$ & $A_{\mathrm{peak}}$ & $A_{\mathrm{map}}$ & $\langle A_{1}\rangle$ & $\langle A_{2}\rangle$ & $A_{\mathrm{vel}}$ \\
        NGC & Mpc$^{-3}$) & & [km\,s$^{-1}$] & [km\,s$^{-1}$] & & & & & $\left(\frac{r}{R_{25}}<1\right)$ & $\left(\frac{r}{R_{25}}>1\right)$ & \\ \hline
        4064 &$-0.44$ & 10.1 & 2 & 3 & 0.442 & 1.05 & 1.29 & 0.28 & 0.09 & --- & 0.110 \\ 
        4189 & 1.06 & 9.6 & 0 & 2 & 0.400 & 1.33 & 1.59 & 0.17 & 0.09 & 0.36 & 0.096 \\ 
        4192 & 0.91 & 10.7 & 13 & --- & 0.560 & 1.17 & 1.35 & 0.19 & 0.39 & --- & 0.079 \\ 
        4216 & 1.03 & 11.0 & 0 & --- & 0.538 & 1.01 & 1.06 & 0.19 & 0.23 & --- & 0.013 \\ 
        4222 & 1.11 & 9.4 & 2 & --- & 1.986 & 1.04 & 1.01 & 0.19 & --- & --- & 0.019 \\ 
        ... & ... & ... & ... & ... & ... & ... & ... & ... & ... & ... & ... \\ \hline
	\end{tabular}
\end{table*}

\begin{table*}
	\centering
    \caption{Asymmetry parameters for HALOGAS galaxies. The full table is available online.}
	\label{table:param_halogas}
	\begin{tabular}{lccccccccccr}
	    \\ \hline
        Galaxy & $\log(\rho_{10}/$ & $\log\left(\frac{M_*}{\mathrm{M}_{\mathrm{\odot}}}\right)$ & $\Delta V_{\mathrm{sys}}$\, & $\Delta V_{\mathrm{sys,opt}}$\, & $A_{\mathrm{spec}}$ & $A_{\mathrm{flux}}$ & $A_{\mathrm{peak}}$ & $A_{\mathrm{map}}$ & $\langle A_{1}\rangle$ & $\langle A_{2}\rangle$ & $A_{\mathrm{vel}}$ \\
        NGC & Mpc$^{-3}$) & & [km\,s$^{-1}$] & [km\,s$^{-1}$] & & & & & $\left(\frac{r}{R_{25}}<1\right)$ & $\left(\frac{r}{R_{25}}>1\right)$ & \\ \hline
        2537 & $-1.42$ & 9.0 & 2 & 17 & 0.202 & 1.20 & 1.27 & 0.15 & 0.02 & 0.27 & 0.044 \\ 
        2541 & $-1.39$ & 8.5 & 2 & 3 & 0.148 & 1.04 & 1.06 & 0.15 & 0.10 & 0.31 & 0.058 \\ 
        3198 & $-0.82$ & 9.6 & 3 & 1 & 0.086 & 1.07 & 1.03 & 0.14 & 0.12 & --- & 0.014 \\ 
        4062 & $-0.72$ & 8.7 & 3 & 1 & 0.092 & 1.08 & 1.06 & 0.05 & 0.09 & 0.41 & 0.010 \\ 
        4244 & $-0.86$ & 10.4 & 2 & 1 & 0.072 & 1.04 & 1.17 & 0.10 & 0.43 & --- & 0.029 \\ 
        ... & ... & ... & ... & ... & ... & ... & ... & ... & ... & ... & ... \\ \hline 
	\end{tabular}
\end{table*}


\bsp	
\label{lastpage}
\end{document}